\documentclass[twocolumn]{aastex631}

\newcommand{\um}{$\mu$m}
\newcommand{\IR}{\textnormal{IR}}

\shortauthors{Chworowsky et al.}
\graphicspath{{./}{figures/}}

\begin{document}

\title{ALMA 1.1mm Observations of a Conservative Sample of High Redshift Massive Quiescent Galaxies in SHELA}

\correspondingauthor{Katherine Chworowsky}
\email{k.chworowsky@utexas.edu}

\author[0000-0003-4922-0613]{Katherine Chworowsky}
\affiliation{Department of Astronomy, The University of Texas at Austin, Austin, TX, USA}

\author[0000-0001-8519-1130]{Steven L. Finkelstein}
\affiliation{Department of Astronomy, The University of Texas at Austin, Austin, TX 78712 USA}

\author[0000-0003-3256-5615]{Justin~S.~Spilker}
\affiliation{Department of Physics and Astronomy, Texas A\&M University, College
Station, TX, 77843-4242 USA}
\affiliation{George P.\ and Cynthia Woods Mitchell Institute for
 Fundamental Physics and Astronomy, Texas A\&M University, College
 Station, TX, 77843-4242 USA}

\author[0000-0002-9393-6507]{Gene C. K. Leung}
\affiliation{Department of Astronomy, The University of Texas at Austin, Austin, TX, USA}

\author[0000-0002-9921-9218]{Micaela B. Bagley}
\affiliation{Department of Astronomy, The University of Texas at Austin, Austin, TX, USA}

\author[0000-0002-0930-6466]{Caitlin M. Casey}
\affiliation{Department of Astronomy, The University of Texas at Austin, Austin, TX, USA}

\author[0000-0001-6842-2371]{Caryl Gronwall}
\affiliation{Department of Astronomy \& Astrophysics, The Pennsylvania State University, University Park, PA 16802, USA}
\affiliation{Institute for Gravitation and the Cosmos, The Pennsylvania State University, University Park, PA 16802, USA}

\author[0000-0002-1590-0568]{Shardha Jogee}
\affiliation{Department of Astronomy, The University of Texas at Austin, Austin, TX, USA}

\author[0000-0003-2366-8858]{Rebecca L. Larson}
\affiliation{Department of Astronomy, The University of Texas at Austin, Austin, TX, USA}

\author[0000-0001-7503-8482]{Casey Papovich}
\affiliation{Department of Physics and Astronomy, Texas A\&M University, College
Station, TX, 77843-4242 USA}
\affiliation{George P.\ and Cynthia Woods Mitchell Institute for
 Fundamental Physics and Astronomy, Texas A\&M University, College
 Station, TX, 77843-4242 USA}

\author[0000-0002-6748-6821]{Rachel S. Somerville}
\affiliation{Center for Computational Astrophysics, Flatiron Institute, 162 5th Avenue, New York, NY, 10010, USA}

\author{Matthew Stevans}
\affiliation{Department of Astronomy, The University of Texas at Austin, Austin, TX, USA}

\author[0000-0002-0784-1852]{Isak G. B. Wold}
\affil{Astrophysics Science Division, Goddard Space Flight Center, Greenbelt, MD 20771, USA}
\affil{Department of Physics, The Catholic University of America, Washington, DC 20064, USA }
\affil{Center for Research and Exploration in Space Science and Technology, NASA/GSFC, Greenbelt, MD 20771}

\author[0000-0003-3466-035X]{{L. Y. Aaron} {Yung}}
\affiliation{Astrophysics Science Division, NASA Goddard Space Flight Center, 8800 Greenbelt Rd, Greenbelt, MD 20771, USA}

\begin{abstract}

We present a sample of 30 massive (log$(M_{\ast}/M_\odot) >11$) $z=3-5$ quiescent galaxies selected from the \textit{Spitzer-}HETDEX Exploratory Large Area (SHELA) Survey and observed at 1.1mm with Atacama Large Millimeter/submillimeter Array (ALMA) Band 6 observations. 
These ALMA observations would detect even modest levels of dust-obscured star-formation, on order of $\sim 20 \ M_\odot \textrm{yr}^{-1}$ at $z\sim4$ at a $1\sigma$ level, allowing us to quantify the amount of contamination from dusty star-forming sources in our quiescent sample. 
Starting with a parent sample of candidate massive quiescent galaxies from the \citet{Stevans21} v1 SHELA catalog, we use the Bayesian \textsc{Bagpipes} spectral energy distribution fitting code to derive robust stellar masses ($M_*$) and star-formation rates (SFRs) for these sources, and select a conservative sample of 36 candidate massive ($M_* > 10^{11}M_\odot$) quiescent galaxies, with specific SFRs at $>2\sigma$ below the \citet{Salmon15} star-forming main sequence at $z\sim4$. 
Based on ALMA imaging, six of these candidate quiescent galaxies have the presence of significant dust-obscured star-formation, thus were removed from our final sample. This implies a $\sim 17\%$ contamination rate from dusty star-forming galaxies with our selection criteria using the v1 SHELA catalog.  This conservatively-selected quiescent galaxy sample at $z=3-5$ will provide excellent targets for future observations to better constrain how massive galaxies can both grow and shut-down their star-formation in a relatively short time period.
\end{abstract}

\keywords{}


\section{Introduction} \label{sec:intro}

Traditionally, it is well established observationally that galaxies follow a strong bimodal color and morphological distribution. Galaxies that are disk dominated and star-forming show strong blue colors, whereas spheroid dominated galaxies that have stopped forming stars and have predominantly old stellar populations are generally red \citep{Baldry2004, Bell2004, Balogh04}. While recent results have pointed to a more nuanced understanding of galaxy colors and populations, and exceptions exist within this bimodality \citep{Fraser_McKelvie_2022, Newman2018}, the majority of galaxy populations still fall within this dichotomy, providing us a powerful tool in characterizing galaxies.
The most massive galaxies in the local Universe are also quiescent, forming stars well below the stellar mass and star-formation relation. The stellar mass and number densities of these galaxies has increased significantly since $z \sim 2$, while the mass density of star-forming galaxies have stayed roughly constant \citep[e.g.][]{Brammer2011, Kriek2006, Fang2012, Muzzin13,  Faber2007, Brennan15}. This implies that there is some physical mechanisms causing star-formation activities to cease, commonly referred to as the process of ``quenching". 

 To attempt to understand the dominant mechanisms responsible for quenching, one can turn to simulations of galaxy formation.  Modern theoretical models are able to reproduce the present day observed quiescent population through a combination of physical effects (e.g. active galactic nuclei (AGN) activity, disk instabilities, mergers). There is now a general consensus that 
 the dominant mechanism leading to the continuation of quenching, suppressing star-formation not only on short timescales but producing galaxies that remain passive through time, is related to feedback from active galactic nuclei (AGN feedback) \citep[e.g.][and references within]{Choi2014, Springel2005a, Fabian_2012, King_2015, Somerville15_annualreview}. 

Historically it has been a challenge for modern theoretical models and simulations to reproduce the observed population of massive quiescent galaxies at higher redshifts ($z>4$), with most models and simulations severely under-predicting the number densities of quiescent sources compared to observations \citep{Brennan15, Merlin18, Merlin2019, ValentinoFrancesco2020QG1B, Cecchi_2019}. 
Constraining when these galaxies appear in the Universe and their abundance at high redshifts can provide critical constraints on the physics responsible for forming and evolving massive galaxies in the early Universe.

Many studies have measured the number densities of quiescent galaxies or the quiescent fraction out to high redshifts using selection criteria based on broadband colors and physical properties derived from photometric spectral energy distribution (SED) fitting \citep[e.g.][]{Dave2019, Muzzin13,Spitler14,Brennan15, Stefanon_2015,Merlin18,Merlin2019, Schreiber18,Glazebrook17, Forrest20, ValentinoFrancesco2020QG1B, Dickey2021}, while recent studies with \textit{JWST} find a significant population of massive quiescent galaxies at $z=3-5$ \citep[][]{Carnall2022}, 
implying that star-formation occurs on faster timescales and quenching more efficiently than predicted by most models.
However, in order to fully understand the timescale of formation for these galaxies, we not only need to identify these galaxies, we must also gather a consensus of their number densities across redshifts. Currently, number densities for massive quiescent galaxy samples at high redshift often differ from one study to another, largely due to differing definitions of quiescence as well as the fact that these surveys have typically probed relatively small volumes contributing to large statistical uncertainties and high field-to-field variance.

\citet{Stevans21} attempted to overcome this statistical uncertainty by performing the current largest volume systematic search for massive quiescent galaxies in the early universe  using deep multi-wavelength imaging across a wide area within the Spitzer/HETDEX Exploratory Large-Area (SHELA) field \citep{Papovich_2016}. 
\citet{Stevans21} presented a photometric catalog, along with a catalog of galaxy properties from SED-fitting for massive galaxies ($M_*/M_\odot > 10^{11}$) at $z=3-5$.  Due to the red nature these galaxies in the optical -- infrared (IR) range, often times optical-IR selected samples of high-redshift massive quiescent galaxies are contaminated by lower redshift dust obscured star-forming galaxies.
Thus, while \citet{Stevans21} initially selected a sample of $\sim$500 quiescent candidate galaxies, in order to limit contamination, they only published an extremely conservative sample of nine candidate massive quiescent galaxies which satisfied all of their cuts even when including a redshift-luminosity prior. This represents a lower limit on the number density in the SHELA field.  For this study we seek to improve upon selection by investigating potential contaminants in the \citet{Stevans21} parent sample, prior to the application of a redshift-luminosity prior.

The somewhat limited depth of the ground-based optical observations mean that many candidate quiescent galaxies have SEDs which are also consistent with dusty star-forming galaxies. 
The presence of dust suppresses emission at UV and optical ranges, while heating of this dust by massive stars results in increased flux at far-IR (FIR) or sub-millimeter wavelengths, allowing us to easily distinguish between star-forming and quiescent galaxies that may look similar in the optical-IR range. 
The deployment of the Atacama Large Millimeter Array (ALMA) allows constraints on the dust emission of the massive galaxy population \citep[e.g.][]{Schreiber17, Casey18, Capak2015, Maiolino2015, Scoville2016}.
By looking at candidate massive quiescent galaxies through ALMA, we can confidently constrain the presence of obscured star-formation across large redshift ranges, thus ruling out dusty star-forming sources in samples \citep{Schreiber18, Santini19}.

In this paper we present the results from ALMA followup of 100 of the original sample of quiescent candidates from \citet{Stevans21}.  We improve upon their selection by deriving galaxy properties from a conservative and more robust SED-fitting re-analysis of these 100 candidates. Using both \texttt{eazy} and \textsc{Bagpipes}, we constrain photometric redshifts and physical parameters of each galaxy, further honing those which are likely to be truly quiescent. We present ALMA Cycle 7 Band 6 1.1 mm observations for these sources to probe for obscured star-formation, ultimately arriving at a higher-confidence and more complete sample of massive quiescent galaxies. 
This paper is organized as follows: in Section \ref{sec:data} we present our parent sample from SHELA. Our method of determining photometric redshifts with \texttt{eazy} and deriving galaxy properties with \textsc{Bagpipes} is described in Section \ref{sec:selecting}. Our ALMA data and measurements are detailed in Section \ref{sec:ALMA}. We discuss the expected contamination of our selection process based on ALMA results in Sections \ref{sec:contamination}  and \ref{sec:discussion}, and we summarize our work and discuss future work in Section \ref{sec:summary}. Where applicable, we assume a cosmology of H$_0$=70 km s$^{-1}$ Mpc$^{-1}$, $\Omega_M$=0.3 and $\Omega_\Lambda$ = 0.7. All magnitudes given are in the
AB system \citep{ABmag}.

\section{Data}\label{sec:data}

We select our sample of massive quiescent galaxies from the SHELA survey.  The SHELA dataset includes modest depth (22.6 AB mag, 50\% completeness) 3.6 \um\ and 4.5 \um\ imaging from Spitzer/IRAC \citep{Papovich_2016}, \textit{u'g'r'i'z'} imaging from the Dark Energy Camera over 18 deg$^2$ (DECam; \citet{Wold19}), VISTA J and K$_s$ photometry from the VICS82 survey (Geach et al. 2017), and a growing database of full-field IFU spectroscopy from Hobby-Eberly Telescope Dark Energy Survey \citep[HETDEX; ][]{Hill08, Gebhardt_2021}, the HETDEX observations in SHELA are presently $\sim$ 20\% complete.  \citet{Stevans21} also obtained imaging with NEWFIRM on the KPNO Mayall 4m with the NEWFIRM HETDEX Survey (NHS; PI Finkelstein), a moderately deep $K_s$ (2.1 \um) near-infrared imaging survey, adding deeper $K_s$=22.4 mag (5$\sigma$) imaging across 22 deg$^2$.  These $K_s$-band data reduce the fraction of catastrophic
errors in photometric redshifts and measure robust
star-formation rates (SFRs) by breaking the age-dust
degeneracy. The SHELA field has also been observed by Herschel/SPIRE at 0.25–0.5mm
(PI Viero; \citet{Viero_2014}), however, these data are very shallow, sensitive to obscured SFRs $ > 200~M_{\odot} \rm{yr}^{-1}$ at $z \sim 1$, and $ > 1000~M_{\odot} \rm{yr}^{-1}$ at $z \sim 3$, much higher than the expected SFRs of the contaminants we search for here, therefore, these data were not included in the analysis presented.

\subsection{Initial Sample}

\citet{Stevans21} constructed a multi-wavelength (0.4-4.5 $\mu$m) $K$-band-selected catalog, using Tractor \citep{Lang2016} to deblend the IRAC photometry. To perform their selection of massive galaxies at $z >$ 3, \citet{Stevans21}  required a $5\sigma$ or greater detection significance in $K_s$, a $2\sigma$ or greater detection in IRAC 3.6 $\mu$m, and a $u'$-band S/N $<2$, as well as a measurement in all optical DECam bands to allow for reliable constraints on $z_\textrm{phot}$. \citet{Stevans21} then used \verb|eazy-py| \footnote{Version 0.2.0-16-g6ab4498; https://github.com/gbrammer/eazy-py}, based on the \texttt{eazy} code \citep{EAZY}, to measure photometric redshifts and stellar population properties for all 1.53 million catalogued galaxies in SHELA. To fit galaxy SEDs, \verb|eazy-py| finds the linear combination of 12 Flexible Stellar Population Synthesis (FSPS) templates (\citet{Conroy09}, \citet{ConroyGunn10}) that minimizes the $\chi^2$ with respect to the fluxes in all available photometric bands. For this initial sample, \citet{Stevans21} ran \verb|eazy-py| with flat priors and selected a parent sample of 3684 massive galaxies (log $(M_*/M_\odot) > 11)$ which have significant detection in $K$ and both IRAC bands with $>60\%$ of their integrated photometric redshift probability density ($P(z)$) at $3<z<5$. Of these galaxies, \citet{Stevans21} found 506 likely quiescent sources, with log(sSFR/yr$^{-1})<-11$.

\citet{Stevans21} also explored the possibility of contamination by red galaxies at low redshift, which can have SED shapes that appear similar to $z=3-5$ quiescent galaxies within our filter set. They attempted to minimize contamination by using an apparent magnitude prior, which applies a Bayesian prior based on a source's apparent magnitude. \texttt{eazy} has a built in prior, derived based on the luminosity functions of the \citet{DeLucia07} semi-analytic model (SAM), which resulted in only nine of the original 506 quiescent candidates being classified as massive quiescent galaxies. \citet{Stevans21} noted that this likely results in significant incompleteness in the published sample. 

\begin{figure}[!t]
\includegraphics[width=0.96\linewidth]{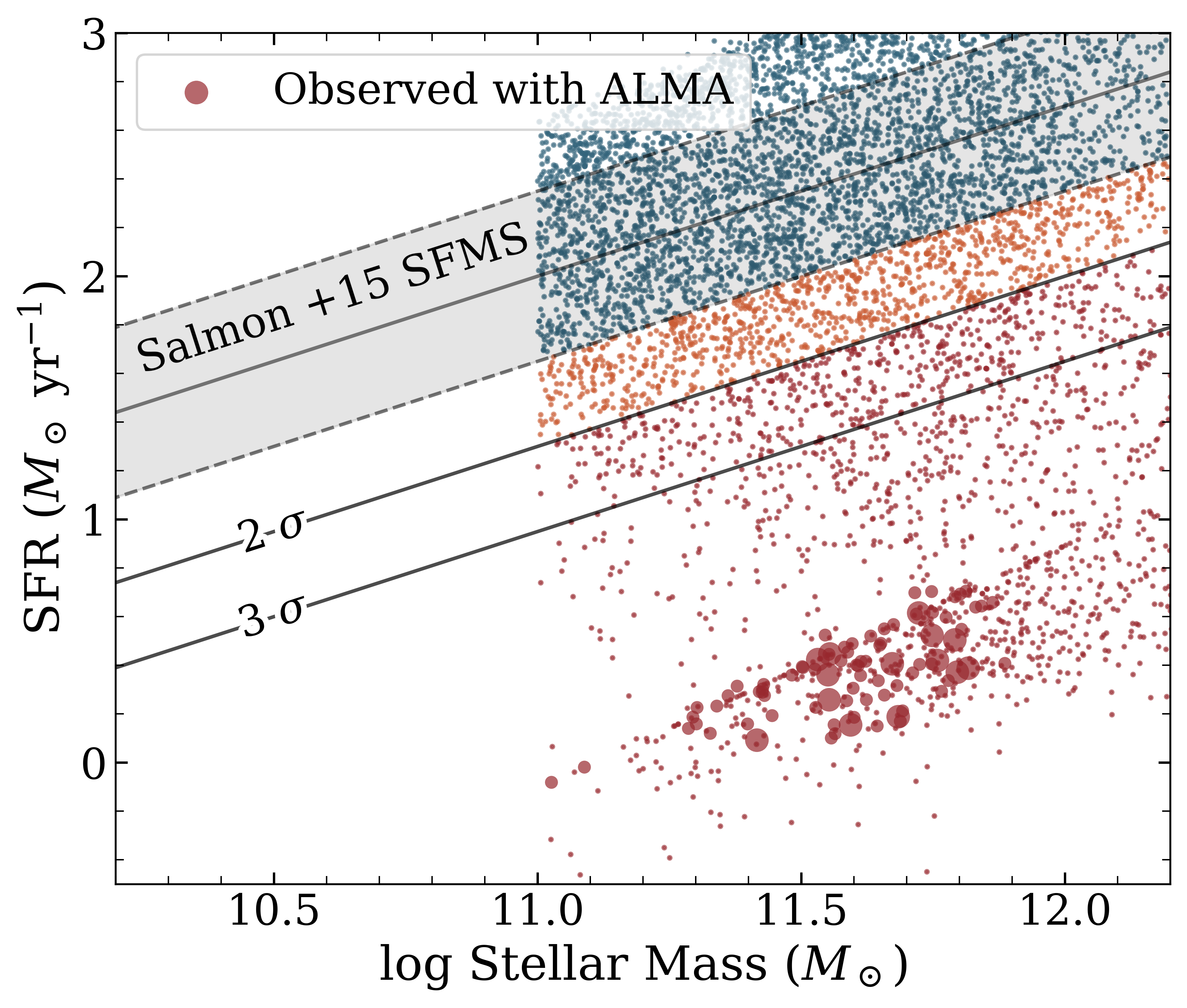}
\caption{The small data points show the $z=3-5$ massive (log$(M_*/ M_\odot) > 11)$ sample from \citet{Stevans21}.  The blue, orange, and red circles denote star-forming, moderately star-forming, and quiescent galaxies, respectively. The red circles are sources with ALMA observations presented here, shown with the \citet{Stevans21} 
\texttt{eazy-py} based SFR and stellar mass values. The larger red circles are $> 3 \sigma$ detections in ALMA.
We also show the \citet{Salmon15} $z =$ 4 star-forming main sequence, with lines highlighting the region of 1, 2 and 3$\sigma$ scatter from the main trend.  
}\label{fig:sample}
\end{figure}

In this study, we chose to revisit their initial parent sample to investigate the contamination by low redshift dusty sources. To achieve this, ALMA Cycle 7 Band 6 1.1 mm observations (rms = 80$\mu$Jy) were obtained for a selected subsample of 100/506 massive quiescent candidates (P.I. Finkelstein). The target sample is selected to span  $3 < z < 5$, log($M/M_{\odot}$) = 11.1 -- 11.9, 20 $< K_s <$ 22, with log(sSFR/yr$^{-1})<-11$ (Figure \ref{fig:sample}). We perform the subsequent analysis on this subset of 100 candidate massive quiescent candidates.

\begin{table*}
\centering
 
\hspace*{-2.5cm}
\begin{tabular}{cccccccc}
\hline
Parameter && Prior && limits \\

\hline
age / Gyr  && Uniform && (0.1, 15) \\

$\mathrm{log_{10}}(M_\mathrm{form}\ /\ \mathrm{M_\odot})$ && Uniform && (1, 15) \\
log$_{10}(U)$ && Uniform && (-4, 2)\\

$Z\ /\ Z_\odot$ && Uniform && (0, 2.5)\\
$\tau \ /\ \mathrm{Gyr}$, timescale of decrease in SFH && Logarithmic && (0.3, 10) \\
$A_V$ && Uniform && (0, 8)\\
$\delta$, deviation from Calzetti slope&& Gaussian$_{\mu=0,\sigma=0.1}$ &&  (-0.3, 0.3) \\
B, 2175 \AA Bump Strength && Uniform && (0, 5) \\

$z_\mathrm{obs}$  && \texttt{eazy} P(z) &&  variable  \\
\hline
\end{tabular}
\centering
 \caption{Fixed and fitted parameters with their associated priors for the declining exponential SFH model and \citet{Salim2018} dust models, assuming a Kroupa initial mass function. $M_\mathrm{form}$ is the total mass of stars formed by the galaxy up to time $\tau$, $\delta$ and B are parameters unique to the Salim 2018 dust models. Priors listed as logarithmic are uniform in log-space.\label{tab:parameters}}
\end{table*}

\section{Selecting Massive Quiescent Galaxies} \label{sec:selecting}

\subsection{Photometric Redshifts with \texttt{eazy}}\label{subsec:EazySED}

While \citet{Stevans21} performed SED fitting using the python version of \texttt{eazy}, \verb|eazy-py|, we elected to perform an independent photometric redshift selection using the more well-tested command-line version of \texttt{eazy}, using the same set of twelve FSPS (\citet{Conroy09},
\citet{ConroyGunn10}) templates (``tweak\_fsps\_QSF\_12\_v3'') as \citet{Stevans21}, which utilize a Chabrier (2003) initial mass function, \citet{Kriek_2013} dust law, and solar metallicity. These FSPS models span
a wide range of galaxy types (star-forming, quiescent, dusty), with different realistic star-formation histories (SFH; bursty and slowly rising). The best-fit combination of templates is
determined through $\chi^2$ minimization. 

As described above, \citet{Stevans21} elected to include a redshift-luminosity prior in their selection of quiescent galaxies. \texttt{eazy} has such a built-in prior, based on the \citet{DeLucia07} SAM. These priors are typically peaked at low-redshift with a tail to higher redshift, with the tail becoming more prominent at fainter magnitudes. This prior will typically make any low-redshift solution be the dominant solution, even if the higher-redshift solution is a better fit to the photometry.
However, the accuracy of luminosity function of galaxies in the \citet{DeLucia07} SAM is in doubt, as the red ($B-V \geq 0.5$) galaxy density at $z \sim 3$ in this SAM under-predicts the observed density by a factor of $\sim 8$ \citep{Marchesini07}. Therefore, we elect to perform our SED fitting without a redshift-luminosity prior to mitigate the possible effects of an unreliable prior.

To select our candidate high redshift sources, we require the integral of the normalized redshift probability distribution function (PDF) to be $>70\%$ for $z>2.5$, compared to \citet{Stevans21} cut of $>60\%$ for $z>3$. Of our initial sample of 100 candidate quiescent galaxies observed with ALMA, 61 satisfied this photometric redshift cut, forming our high-redshift sample.

\begin{figure*}[!t]
\includegraphics[width=0.93\linewidth]{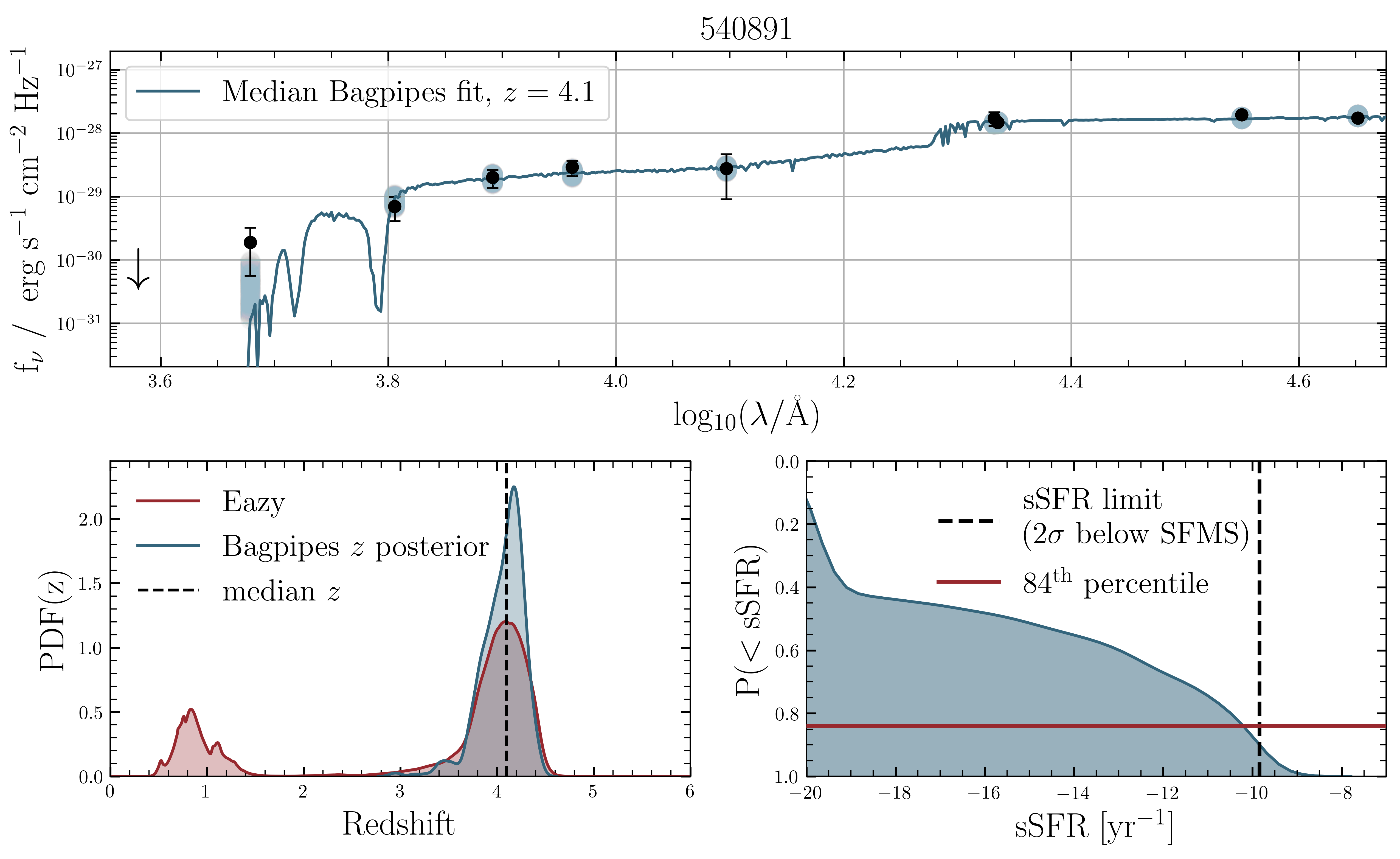}
\centering
\caption{An example of \textsc{Bagpipes} results for a single source. The median SED is plotted in the top panel. The black points show the observed fluxes from SHELA and the blue shaded points are the 1$\sigma$ range of the posterior photometry. The lower left plot shows the redshift distribution from \texttt{eazy} in red, and the posterior redshift distribution from \textsc{Bagpipes} in blue. The bottom right panel shows the posterior on the sSFR: the black dashed line shows the sSFR limit for our quiescent selection, and the red line shows the median value of the posterior in sSFR. \label{fig:BP_fullfit}}
\end{figure*}

\subsection{SED Fitting with \textsc{Bagpipes}}\label{subsec:Bagpipes}
While \texttt{eazy} provides a robust photometric redshift determination, due to the small set of templates, it has a limited ability to explore the full parameter space of physical properties of galaxies. Thus, in order to determine the stellar mass and star formation rates of our sources, we utilize
the \textsc{Bagpipes} modeling code. \textsc{Bagpipes} is a Bayesian spectral fitting code which models the emission from galaxies from the far-ultraviolet to the millimeter regime, allowing the user to build up complex models for fitting, with user-defined stellar population synthesis models, star formation histories, and dust attenuation and emission models. \textsc{Bagpipes} is a Python tool, and can fit these models to arbitrary combinations of spectroscopic and photometric data using the \textsc{MultiNest} nested sampling algorithm \citep{Carnall18}. 

After using \texttt{eazy} to generate our sample of high redshift sources, we run \textsc{Bagpipes} on the SHELA observations. We modify \textsc{Bagpipes} to take as input the redshift probability distribution as determined by \texttt{eazy} as a prior on the modelled redshift. 
We further force \textsc{Bagpipes} to ignore any potential low-redshift solution (already limited to $<$30\% of the redshift probability density) by setting P($z <$ 2.5) to zero. This allows us to assume that the high-redshift solution is generally correct,  essentially implementing an additional prior that each source is at $z>2$, but also to generate a posterior probability distribution of the stellar masses and SFRs of each source with uncertainties inclusive of the photometric redshift uncertainties. We elected to include the photometric redshift prior since \texttt{eazy} is optimized for photometric redshift recovery, whereas \textsc{Bagpipes} is allowed to explore a large range of parameter space. Furthermore, allowing for lower redshift solutions would result in unphysical marginalized uncertainties on the physical parameters of interest, which would combine both low- and high-redshift solutions.  Therefore, by constraining the redshift space, we force \textsc{Bagpipes} to explore a wider range of galaxy parameters within the likely redshift of each source.

As our parent sample was initially selected to be quiescent, each galaxy was fit with a declining exponential star-formation history with a logarithmic prior on $\tau$, the timescale of star formation decline; the complete fit parameters are listed in Table \ref{tab:parameters}. We also compared \textsc{Bagpipes} ability to probe the complete parameter space by changing the number of live points (\verb|n_live|) in the MCMC fit and found that a value of 3000 live points produces the best fit results within a reasonable fit time. Each SED fit is returned with a posterior distribution in mass and SFR (an example is shown in Figure \ref{fig:BP_fullfit}), from which we classify sources as quiescent based on the distribution of the sSFR posterior. SEDs for the entire high redshift sample can be seen in Figures \ref{fig:BP_SEDs_noALMA} and \ref{fig:BP_SEDs_ALMA} in the Appendix.

\begin{figure}[!t]
\centering
\includegraphics[width=0.99\linewidth]{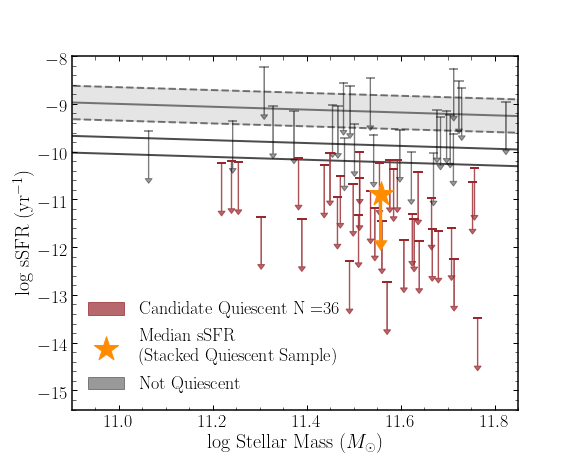}
\caption{The sSFR vs stellar mass distribution of our 61  high redshift sources. The grey shaded region is the $1\sigma$ range of the \citet{Salmon15} SFMS, the grey solid lines are 2 and 3 $\sigma$ below the SFMS. To be classified as quiescent, we require 84\% of the integral of the P(sSFR) from \textsc{Bagpipes} to fall below 2$\sigma$ of the SFMS for the given best-fit stellar mass (red). For the sources not selected as quiescent (grey arrows), the majority have a sSFR median well below the SFMS at the best-fit stellar mass.  However, their sSFR distributions are broad, such that the 84\% of the integral of the P(sSFR) falls above our 2$\sigma$ selection. For the 36 quiescent candidates shown in red, we plot the upper limits from our \textsc{Bagpipes} fit if the median of the SFR is less than 0.1 $M_\odot$yr$^{-1}$, which is the case for all sources. The yellow star shows the median stellar mass and sSFR measured from SHELA for the sample of 36 candidate massive quiescent galaxies. \label{fig:select_final}}
\end{figure}

To determine which of our sources are likely quiescent, we compare the derived specific star formation rates to the $z =$ 4 star-forming main sequence (SFMS) as presented in \citet{Salmon15}. To be classified as quiescent, we require 84\% of the sSFR posterior to be $>2\sigma$  below the SFMS at the best-fit photometric redshift. We note that this is a much more conservative selection than the typical process of using best-fit values.  In fact, while the majority of our galaxies have a sSFR median well below the SFMS at the best-fit stellar mass, this method allows us to encompass the full uncertainty on the sSFR posterior due to other possible solutions in our SED modeling.  We also note the difference from the \citet{Stevans21} selection, who used a fixed sSFR threshold of log (sSFR/yr$^{-1}) < -$11.  By using the $z =$ 4 main-sequence as a threshold, our selection accounts for the evolution of galaxy typical sSFRs to higher values with increasing redshift \citep{Speagle2014, Rinaldi2022, Sandles_2022}.  With this conservative selection, we find that 36 out of the 61 high redshift sources were classified as quiescent based on optical-IR SED fitting. These sources are shown in Figure \ref{fig:select_final}.  To summarize, based on our updated photometric redshift and stellar population modeling, we consider 36 of the original 100 ALMA-observed galaxies from \citet{Stevans21} to be robust candidate massive quiescent $z >$ 3 galaxies, with the smaller number due to our more advanced selection procedure. 

\section{ALMA Observations} \label{sec:ALMA}

\begin{figure}[!t]
\centering
\includegraphics[width=0.99\linewidth]{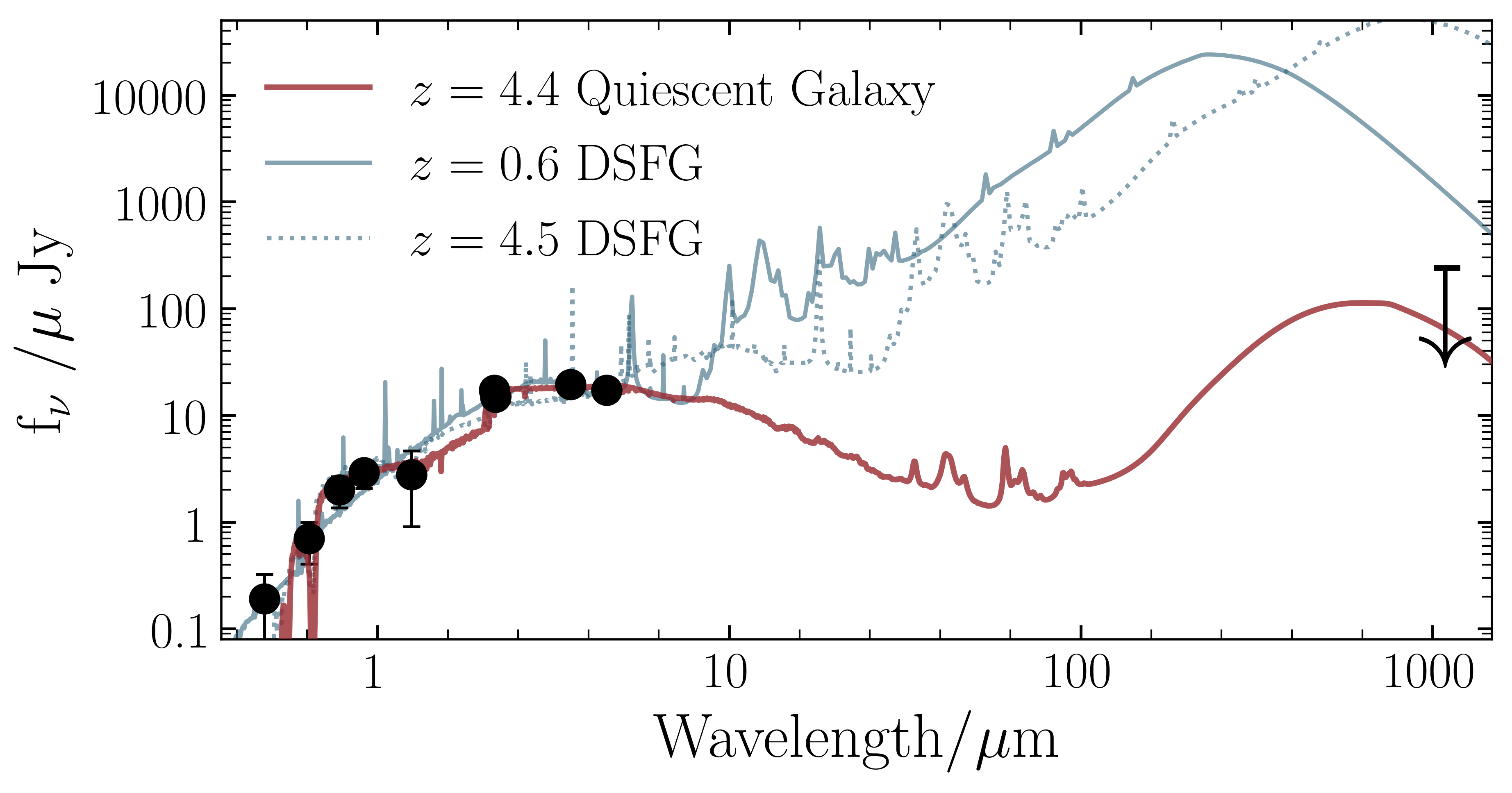}
\caption{Example photometry of one of our quiescent sources (black) with the best-fit \textsc{Bagpipes} SED and example dusty star-forming galaxy (DSFG) SEDs at both low and high redshifts. Although the SHELA photometry is best-fit to a high redshift quiescent galaxy (red), with sSFR=-10.3, it is also consistent with high redshift DSFG with sSFR=-8.0 (dotted blue) and a low redshift DSFG solution with sSFR=-8.5 (solid blue). The ALMA 3$\sigma$ depth is shown in the downward arrow, significantly ruling out dust obscured star-formation at any redshift. \label{fig:SED_comp}}
\end{figure}

For the sample of galaxies presented here, we have obtained ALMA Band 6 1.1mm observations in Cycle 7 program 2019.1.01219.S (PI Finkelstein).  These data were obtained in December 2019.
Each source was observed 6 $\times$ 26.4 seconds, for a total integration time of 158 seconds for each source. While relatively short integrations, the sensitivity of ALMA allows us to reach  sufficiently deep to robustly detect dust emission from contaminating dusty star-forming galaxies (and possibly detect weak low-level star formation from truly quiescent galaxies). Figure \ref{fig:SED_comp} illustrates how the SEDs of quiescent and dusty star-forming galaxies diverge at longer wavelengths, rest-frame far-IR (FIR) observations to a sensitivity of 80 $\mu$Jy at 1.1mm with ALMA can robustly rule out star-forming solutions for our quiescent candidates.

We use the default continuum images of each source produced by the ALMA pipeline in our analysis. These images use Briggs weighting with robustness parameter 0.5, and reach a typical spatial resolution of 1.30" $\times$ 0.80" (corresponding to a physical size of order 9  $\times$ 6 kpc at $z=4$). 
For quiescent sources, We do not expect these galaxies to be spatially resolved in our ALMA observations, as high redshift quiescent galaxies are often compact, with half-light sizes of a few kpc \citep{Straatman_2015, Stefanon13}.
For sources with possible IR-emission (i.e. dust obscured star-forming galaxies), previous studies with sufficient resolution in ALMA to measure sizes suggest that dust emission takes place within compact regions of the galaxy, with observed $R_e$(FIR) to be generally smaller than $R_e$(optical), on order of $\lesssim 2$ kpc at $z \sim 3$, \citep{Fujimoto_2017, Franco18, Ikarashi2017}. In both cases, quiescent or dusty star-forming galaxies, the galaxy is expected to be roughly point-like for the spatial resolution of the presented observations. 

\begin{figure}[ht]
\centering
\includegraphics[width=0.85\linewidth]{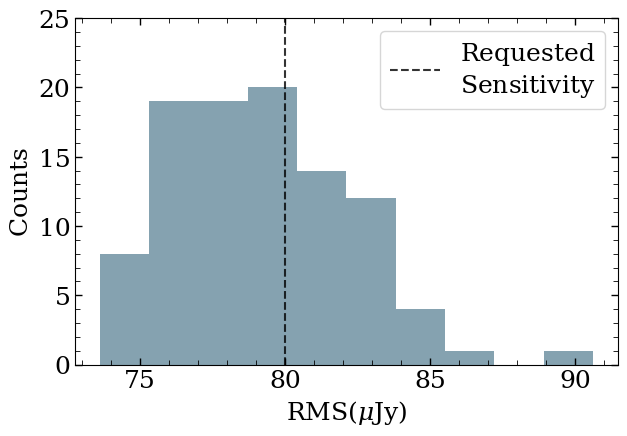}
\caption{The distribution of the rms in our 98 ALMA observations. The reported rms values are calculated by removing pixels with $>$5$\sigma$ detections, and then calculating the rms of the remaining pixels. The median rms is 79 $\mu$Jy, with a standard deviation of 3 $\mu$Jy, comparable to our desired sensitivity of 80 $\mu$Jy.\label{fig:ALMA_rms}}
\end{figure}

From the 100 targets observed, we visually examined each source observation and removed 2 due to observational artifacts, resulting in 98 sources with reliable data. We measured the sensitivity in each observation as the 5$\sigma$ clipped rms of the images from the non primary-beam corrected map.  We show the distribution from our 98 observations in Figure \ref{fig:ALMA_rms}.  The median value of all images is 79 $\pm$ 3 $\mu$Jy/beam. Of the 36 robust candidate high redshift massive quiescent galaxies, all had reliable data, and we perform the following analysis on this subsample. We also present results from the 62 sources not included in our subsample in Appendix \ref{sec:not_selected}.

\subsection{Astrometric Corrections} \label{subsec:astrometry}
 To determine the source emission, we first match the source location between SHELA and the ALMA observations. The DECam exposures in SHELA were observed through a seven-year period, and the images were reduced using different versions of the NOAO DECam Community Pipeline, therefore, we performed a re-calibration of the astrometry and flux scaling of each image uniformly prior to stacking. The process is summarized here, but details can be found in the updated SHELA catalog paper \citep{Leung23}. The astrometry of each image is tied to the Gaia EDR3 catalog \citep{gai21}. For each image, an initial source catalog is generated using \texttt{SEP} \citep{Barbary2016} and matched to the reported coordinates of astrometry stars in the Gaia EDR3 catalog. The median $x$- and $y$-offset required to match the good stars to the Gaia coordinates in each image was determined and applied to SHELA to correct the astrometry. We use these astrometric corrected coordinates as the source position in our measurements on the ALMA band 6 imaging.

\subsection{Source Flux Determination}\label{subsec:ALMAflux}

In order to determine the integrated flux of each ALMA observed source, we employed the Common Astronomy Software Applications package (\textsc{CASA}) \verb|imfit|\footnote{https://casa.nrao.edu/docs/taskref/imfit-task.html} task, which attempts to fit one or more elliptical Gaussian components on an image region. We performed two iterations of \verb|imfit|.  For the first iteration, we fix the elliptical Gaussian to the ALMA synthesized beam shape with the center of the beam allowed to move freely (beam-fixed). This method allows us to recover accurate fluxes accounting for slight deviations of the source position due to astrometry in the SHELA catalog. In the second iteration, we fix both the elliptical Gaussian to the ALMA synthesized beam shape, and the center to the Right Ascension and Declination of the source (beam$+$position-fixed), accounting for non-detections. The initial estimates provided to \verb|imfit| are chosen to be the astrometrically corrected center of the quiescent candidate and the flux of the pixel at that location, since the sources of interest are unresolved by ALMA. 

We use the integrated flux as determined from the beam-fixed fit as our fiducial flux value if the fit returns a source with S/N $>$ 3.  Our tests show that at that significance level, \verb|imfit| is able to accurately find the source center. Four sources (out of the full selected parent sample of 36) were found to have a source with S/N $>$ 3, and thus were fit in this way. Within these images, we then determine the separation of the \verb|imfit| determined source from the expected source RA and Dec. We determine that if the fitted center is within the \textit{K}-band PSF (FWHM $\sim1.41"$), that the emission is indeed coming from the quiescent candidate. All four images with S/N $>3$ were determined to be consistent with the quiescent candidates from the parent sample.  
For the sources for which the beam-fixed fit found a S/N $<$ 3, we chose to use the integrated flux as calculated from the beam+position-fixed fit as the flux of our source, since the reliability of the fitted position for a source with such low S/N is questionable. The remaining 32 candidate massive quiescent galaxies were all found to have a S/N $<$ 3 at the center, therefore we chose to report the flux calculated from the beam+position-fixed fits. 

\section{Obscured Star-Formation Rates \& Contamination} \label{sec:contamination}

ALMA 1.1mm observations measures dust emission corresponding to dust obscured star-formation, allowing us to remove obvious contaminants from our quiescent sample.  In order to determine the obscured SFR of each source, we generate model dust SEDs at our ALMA sensitivity limit based on the prescription of \citet{Drew_2022}: we adopt a range of dust temperatures and fit the ALMA 1.1mm flux to a modified optically thin black-body with a mid-infrared powerlaw of index $\alpha \sim 2$. We take three sample SEDs at the wavelength of the peak dust temperature ($T$) and the upper and lower 1$\sigma$ range, based on our ALMA sensitivity. For the quiescent candidates, this corresponds to $T$ = 25$\pm$ 5K.
We then scale each dust SED to the ALMA observations to calculate each galaxies' infrared luminosity ($L_{\textrm{IR}}$) by integrating each scaled dust SED from $8-1000 \mu$m and assuming each source is at the best fit redshift (also the peak of $P(z)$, $z_a$) from \texttt{eazy}.
From this $L_{\textrm{IR}}$, we then 
approximate the level of obscured star-formation following \citet{Kennicutt_2012}, which assumes the \citet{Kroupa2003} IMF with a Salpeter slope of $\alpha_*=-2.35$ from $1-100M_\odot$ and $\alpha_*=-13$ from $0.1-1M_\odot$:
\begin{equation}
    \textrm{SFR} =\textrm{log} (L_{\IR}) - 43.41
\end{equation}
Where SFR is given in $M_\odot \ \textrm{yr}^{-1}$ and $L_{\textrm{IR}}$ is in $\textrm{erg} \ \textrm{s}^{-1}$. 

To determine $L_\textrm{IR}$, each of the 36 sources is fit with a 20, 25, and 30 K (characteristic dust temperature $T$ = 25$\pm$ 5K) modified black-body as described above. For these dust temperatures, a 1$\sigma$ detection ALMA 1.1mm flux at our sensitivity limit ($\sim 80 \mu$Jy) would result in an obscured SFR upper limit of $\sim$ 15, 21, 27  $M_\odot \ \textrm{yr}^{-1}$, respectively. 

\begin{figure*}[!hb]
\includegraphics[width=0.95\linewidth]{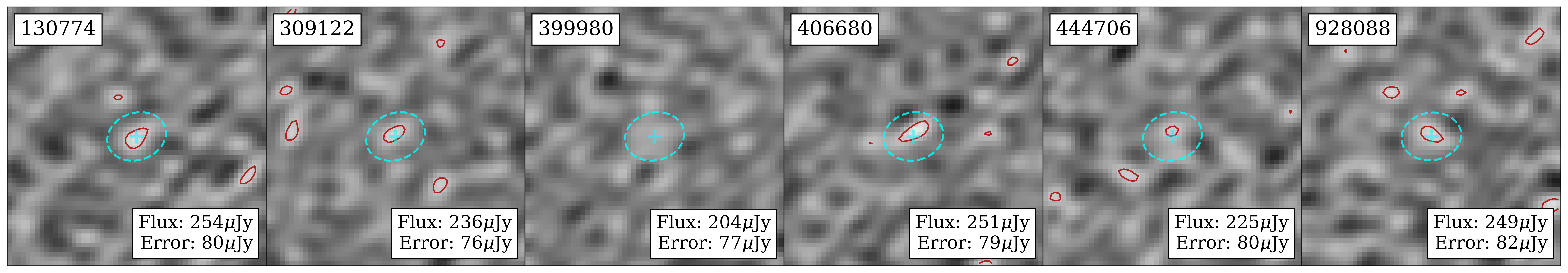}
\caption{ALMA 1.1 mm images for the 6 of our 36 quiescent candidates for which $>1\sigma$ ALMA flux was detected and, based on estimated obscured SFR, were removed from our quiescent sample, we note that, all of these sources in fact have a S/N$>2.5$ in the ALMA Band 6. The red lines are 3$\sigma$ contours, the source RA and Dec are denoted by the cyan cross and the fitted ALMA beam is shown as the cyan ellipse. Each stamp is 10"$\times$10".\label{fig:dust_detected}}
\end{figure*}

Six sources that satisfy our massive quiescent selection have measured dust-obscured SFRs that place them above our sSFR selection threshold using these characteristic dust temperatures. We thus refit these sources using dust models more typical of dusty star-forming galaxies at $z\sim 4$, with $T$=40$\pm$5 \citep{Sommovigo_2022} and recalculate the dust-obscured SFRs. The ALMA Band 6 images of these sources are shown in Figure \ref{fig:dust_detected}, the ALMA fluxes as well as estimated obscured SFRs for these sources are reported in Table \ref{tab:ALMA_detected}. We note that detections in ALMA do not necessarily imply that these are not high redshift sources, instead it only speaks to the existence of significant star-formation activity, therefore are no longer characterized as quiescent in this conservative sample.

\begin{figure}[h]
\centering
\includegraphics[width=0.7\linewidth]{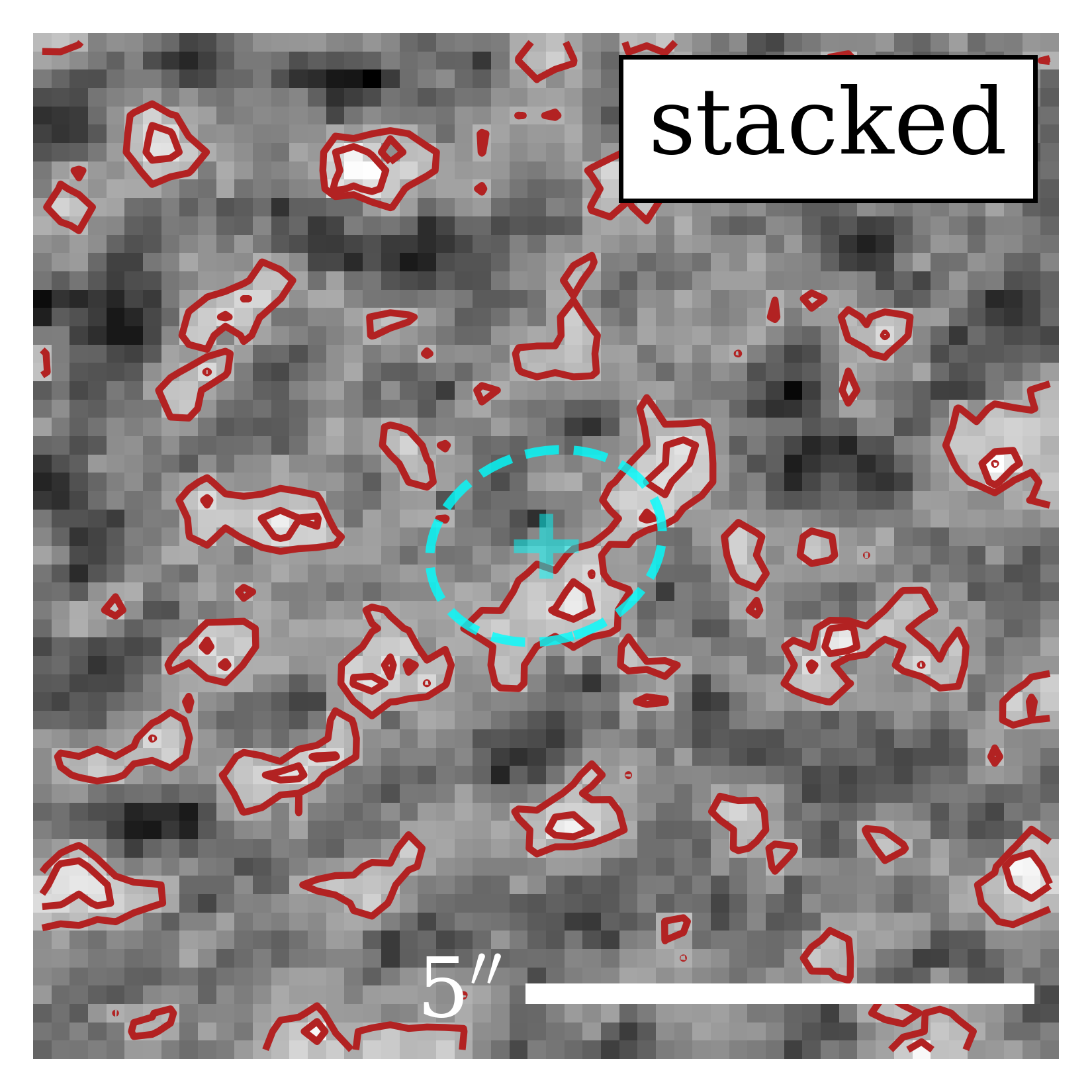}
\caption{The median stack of our final sample of 30 massive quiescent galaxies. The cyan ellipse is the median ALMA beam centered on the expected source positions. The red denotes 1- and 2-$\sigma$ contours, the measured flux is 9$\mu$Jy  $\pm 17 \mu$Jy. \label{fig:stacked}}
\end{figure}

 \subsection{Quiescent Sample}\label{subsec:quiescent_sample}
\begin{figure*}[hb]
\includegraphics[width=0.95\linewidth]{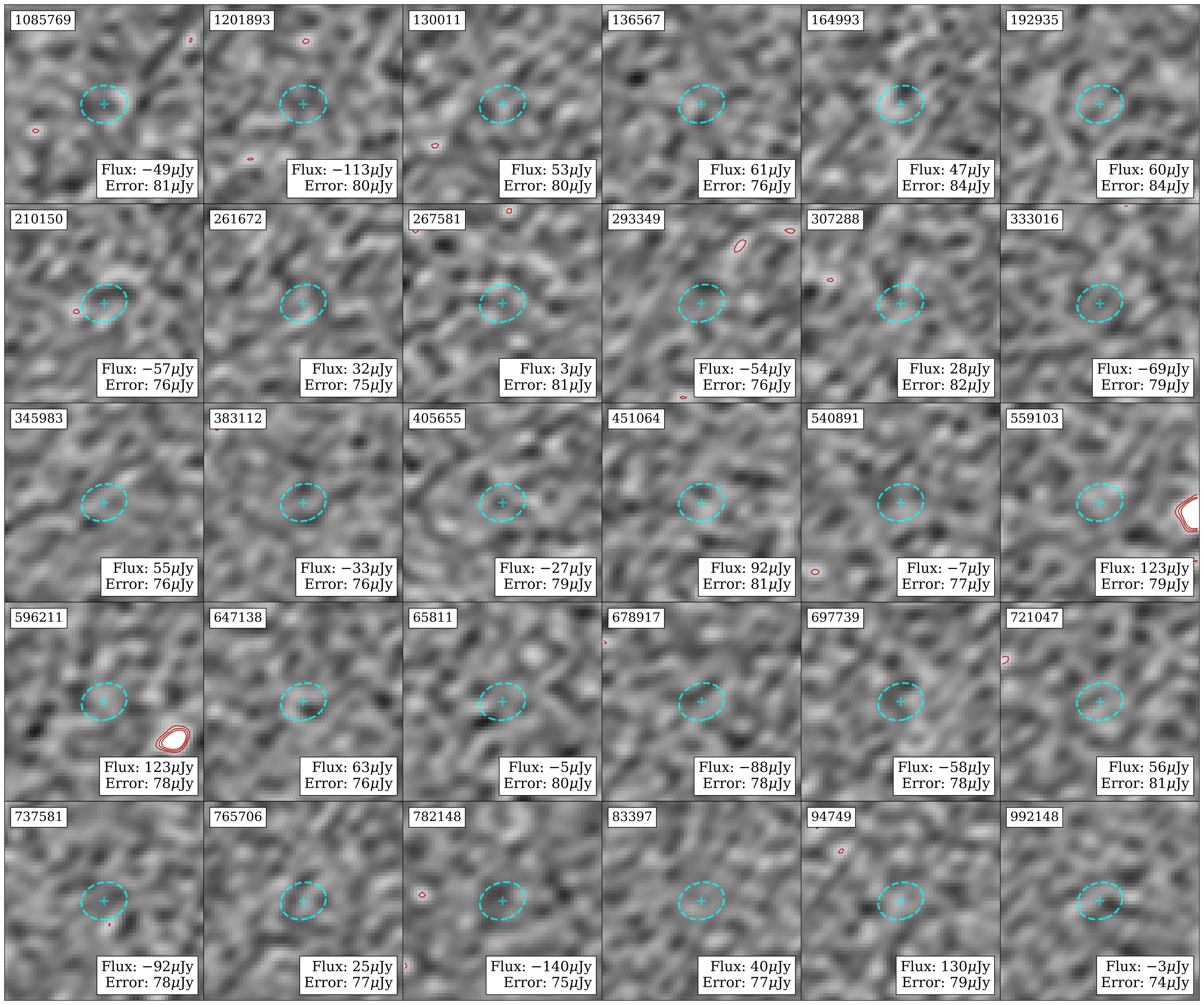}
\caption{ALMA images for the 30 selected quiescent galaxies. The red lines are 3, 4, and 5 $\sigma$ contours, the ALMA beam is shown as the cyan ellipse, and the source RA and Dec is denoted by the cyan cross. To be characterized as quiescent, we calculate the dust-obscured SFR from the measured flux and require the upper 1$\sigma$ limit of the SFR to fall 2$\sigma$ below the SFMS at the galaxies best-fitting stellar mass. Each stamp is 10"$\times$10". \label{fig:ALMA_srcs1}}
\end{figure*}

\begin{figure}[h]
\centering
\includegraphics[width=0.99\linewidth]{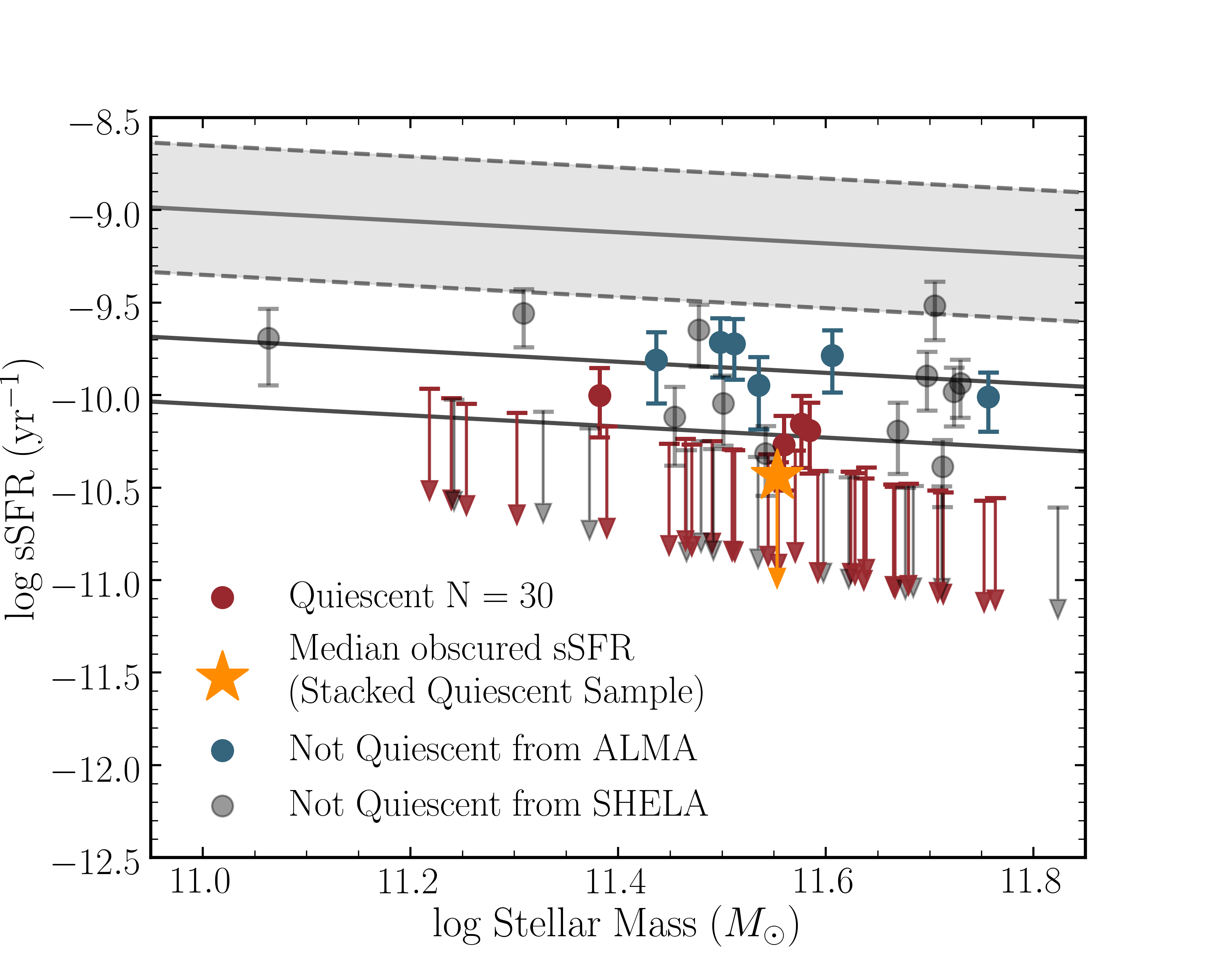}
\caption{Similar to Figure \ref{fig:select_final}, but shown here are the measured sSFR based on ALMA 1.1mm imaging. The grey shaded region is the $1\sigma$ range of the \citet{Salmon15} SFMS, the grey solid lines are 2 and 3 $\sigma$ below the SFMS. The distribution of our 61 high redshift sources with respect to the star-forming main sequence. The grey points are sources removed from our quiescent sample from SED fitting with \textsc{Bagpipes} (\S \ref{subsec:Bagpipes}). For the 36/61 sources classified as quiescent from optical data (colored points), we then examine ALMA band 6 1.1mm observations.
Similar to the measured optical SFR, to be classified as quiescent, we
require the upper $1\sigma$ limit of the FIR SFR
to fall below $2\sigma$ of the SFMS for the given best-fit stellar mass as recovered by \textsc{Bagpipes}. This removed 6/36 sources (shown in blue) from our final quiescent sample (red,details in \S \ref{sec:contamination}).\label{fig:SFMS_ALMA}}
\end{figure}

The removal of the six sources with significant obscured SFRs from our quiescent sample leaves us with a final sample of 30 candidate massive quiescent galaxies, images for these sources are shown in Figure \ref{fig:ALMA_srcs1}, the reported errors are the 5$\sigma$-clipped rms of the images. This implies that we expect a contamination rate of 6/36, or $\sim$ 17\%, based on SED fitting using only SHELA photometry for selecting massive high redshift quiescent galaxies. 
To explore the constraints on the obscured SFR made possible from the ALMA imaging, we create a median stack of the 30 quiescent galaxies, shown in Figure \ref{fig:stacked}. We show the distribution of the measured dust-obscured star formation with respect to the SFMS of our sample in Figure \ref{fig:SFMS_ALMA}. We determine the median flux of our final quiescent sample by repeating the analysis detailed in \S \ref{subsec:ALMAflux} using a beam+position fixed \verb|imfit| fit on the stacked image, finding a flux density of 9 $\pm$ 17 $\mu$Jy, a $<1 \sigma$ detection.

We thus calculate the corresponding obscured SFR at the median redshift of our sample in the same way as above, using the 1$\sigma$ flux error of 17 $\mu$Jy as the flux, and find an obscured SFR upper limit of $<$ 4  $M_\odot$/yr$^{-1}$. At the median mass of our sample of  log($M/M_{\odot}$) = 11.5, this corresponds to a sSFR of $-10.9$ yr$^{-1}$. We show the median stellar mass and upper limit in obscured SFRs of our final sample in Figure \ref{fig:SFMS_ALMA}. 

\begin{table*}
\centering
\hspace*{-2.5cm}
\begin{tabular}{cccccccccc}
\hline
SHELA ID & $z_a$ &&  K-band Mag & ALMA 1.1mm flux [$ \mu \textrm{Jy}$] && Obscured SFR  [$ M_\odot \textrm{yr}^{-1}$] & log sSFR [$\textrm{yr}^{-1}$] \\
\hline
406680 & 3.2  && 21.2 & 251  $\pm$ 79 
&& 66  $\pm$ 25 & -9.8 $\pm$ 0.2\\
130774 &3.2 && 21.1 & 254 $\pm$ 80 
 && 60  $\pm$ 21 & -9.7 $\pm$ 0.1\\
309122 & 3.1  && 20.0 & 236  $\pm$ 76 
 && 56  $\pm$ 20 & -10.0 $\pm$ 0.2\\
928088 & 3.1 && 20.8 & 249 $\pm$ 82 
&& 62  $\pm$ 22 & -9.7 $\pm$ 0.2\\
399980 & 4.2 && 20.8 & 204 $\pm$ 77 
&& 46 $\pm$ 16 & -9.8 $\pm$ 0.1 \\
444706 & 4.5 && 21.3 & 225 $\pm$ 80
&& 50 $\pm$ 18 & -9.7 $\pm$ 0.2\\
\hline 
\end{tabular}
\centering
 \caption{ALMA fluxes and obscured SFRs of the six sources from our initial sample of 36 candidate quiescent galaxies removed based on flux measurements in AMA 1.1mm imaging, assuming these sources are at the \texttt{eazy} photometrically-fitted redshifts. These galaxies were classified as quiescent based on SED fitting of SHELA photometry, however, flux in ALMA 1.1mm indicate the presence of obscured star-formation, therefore were removed from the final sample of quiescent sources. We report the estimated dust-obscured SFRs calculated from a dust model with characteristic dust temperature $T$=45$\pm$5K expected from dusty star-forming galaxies at $z \sim 4$. Based on these results, we expect a contamination rate of $\sim $ 17\% in future quiescent samples within SHELA. We report the obscured SFR as described in Section \ref{sec:contamination}.}\label{tab:ALMA_detected}
\end{table*}

\section{Discussion} \label{sec:discussion}

 This work presents a robust exploration into the contamination rate in a sample of $z >$ 3 massive quiescent galaxy candidates. The primary contaminants of massive quiescent galaxies at $z>3$, particularly in the optical-IR range, are lower redshift ($z \sim 1-2$) dust obscured star-forming galaxies. However, these galaxies will be easily distinguishable at longer wavelengths as this dust will be detected in emission with submm/mm observatories such as ALMA. 
 
 In this work, we develop a methodology to robustly select high-redshift ($z >$ 3) massive quiescent galaxies from broadband imaging (in the SHELA field). We show that by examining ALMA observations for our samples of candidate high redshift quiescent galaxies only 6/36 sources have significant ALMA flux.  This implies a contamination rate of $\sim 17\%$ from dust obscured star-forming galaxies with SFRs of $\gtrsim 20 \ M_\odot/\textrm{yr}^{-1}$ (calculated from a 1$\sigma$ detection of 80 $\mu$Jy, corresponding to a log (sSFR/yr$^{-1})<-10.2$ at log($M_*/M_\odot)=11.5$).  This relatively low contamination rate indicates that future work in the SHELA field using these selection criteria can compose a robust sample of massive quiescent galaxies with limited contamination from dusty star-forming galaxies.  This lays the groundwork to leverage the extraordinary volume made available by SHELA to measure robust number densities of these sources across $z=3-5$.

\section{Summary \& Future Work} \label{sec:summary}

We present a robust selection of high redshift massive quiescent galaxies based on photometric fitting, with a goal of exploring contamination via ALMA Band 6 1.1 mm imaging.  We begin with a parent sample of 100 sources in the SHELA catalog, fit their photometric redshifts using \texttt{eazy}, and limit ourselves to a sample of 61 high redshift sources using a conservative photometric-redshift selection criteria of 70\% of the redshift probability integral $P(z)>2.5$. 

From these sources, we perform SED fitting using \textsc{Bagpipes} to constrain the physical properties of each source, and select likely quiescent sources. We found that 36/61 high redshift sources were quiescent (with SFR $>2\sigma$ below the SFMS from \citet{Salmon15} at $z=4$). We then examine ALMA Band 6 1.1mm imaging, which allows us to rule out significant dust obscured star-formation activity and found 6/36 sources were likely star-forming. We determined the obscured SFR by fitting model dust SEDs at our ALMA sensitivity limit based on the prescription of \citet{Drew_2022} to obtain a $L_{\textrm{IR}}$. We then approximate the level of obscured star-formation following \citet{Kennicutt_2012}. We report the sSFR of these six sources, confirming that they lie above our quiescent threshold, and removed them from our final sample of massive quiescent galaxies.  These results imply that the selection of massive quiescent sources based on SED fitting of the SHELA catalog will have an expected contamination rate of $\sim$ 17\% by dusty star-forming galaxies.

This results in a final sample of 30 massive (log$(M/M_\odot)>11$) quiescent galaxies at $z>3$.  It is important to note that this is a preliminary sample, the goal of this paper is to determine a selection criteria and measure the contamination by dusty star-forming galaxies by utilizing ALMA observations.
We intend to repeat this SED fitting process with the updated SHELA catalog \citep{Leung23}, which will be 1-1.3 mag deeper across all bands, as well as include newly acquired \textit{Y}-band imaging to better constrain our future sample. 

This sample of 30 massive high redshift quiescent galaxies also provide excellent potential targets for follow-up spectroscopy with \textit{JWST} NIRSpec spectroscopy. Deep spectroscopy from \textit{JWST} will be able to detect Balmer absorption features at these redshifts and provide the all-important redshift confirmation for these extreme sources. With NIRSpec, observations of key star-formation features (H-alpha emission, Balmer absorption, D$_n$4000, UV/IR slopes) will also allow investigations into the details of the formation and subsequent quenching of high redshift massive quiescent galaxies, which are encoded in their stellar populations. 

\begin{acknowledgments}
KC, SLF and GCKL acknowledge support from the National Science Foundation through grant AST-2009905. This material is based upon work supported by the National Science Foundation Graduate Research Fellowship under Grant No. DGE 2137420. This paper makes use of the following ALMA data: ADS/JAO.ALMA\#2019.1.01219.S. ALMA is a partnership of ESO (representing its member states), NSF (USA) and NINS (Japan), together with NRC (Canada), MOST and ASIAA (Taiwan), and KASI (Republic of Korea), in cooperation with the Republic of Chile. The Joint ALMA Observatory is operated by ESO, AUI/NRAO and NAOJ. The National Radio Astronomy Observatory is a facility of the National Science Foundation operated under cooperative agreement by Associated Universities, Inc. AY is supported by an appointment to the NASA Postdoctoral Program (NPP) at NASA Goddard Space Flight Center, administered by Oak Ridge Associated Universities under contract with NASA.
We thank Adam Carnall for helpful conversations about the running of Bagpipes.
\end{acknowledgments}

%

\vspace{5mm}

\software{Astropy, \texttt{Eazy}, \textsc{Bagpipes}, \texttt{CASA}}




\bibliography{ALMA_quiescent}{}

\begin{thebibliography}{}
\expandafter\ifx\csname natexlab\endcsname\relax\def\natexlab#1{#1}\fi
\providecommand{\url}[1]{\href{#1}{#1}}
\providecommand{\dodoi}[1]{doi:~\href{http://doi.org/#1}{\nolinkurl{#1}}}
\providecommand{\doeprint}[1]{\href{http://ascl.net/#1}{\nolinkurl{http://ascl.net/#1}}}
\providecommand{\doarXiv}[1]{\href{https://arxiv.org/abs/#1}{\nolinkurl{https://arxiv.org/abs/#1}}}

\bibitem[{{Baldry} {et~al.}(2004){Baldry}, {Glazebrook}, {Brinkmann},
  {Ivezi{\'c}}, {Lupton}, {Nichol}, \& {Szalay}}]{Baldry2004}
{Baldry}, I.~K., {Glazebrook}, K., {Brinkmann}, J., {et~al.} 2004, \apj, 600,
  681, \dodoi{10.1086/380092}

\bibitem[{{Balogh} {et~al.}(2004){Balogh}, {Baldry}, {Nichol}, {Miller},
  {Bower}, \& {Glazebrook}}]{Balogh04}
{Balogh}, M.~L., {Baldry}, I.~K., {Nichol}, R., {et~al.} 2004, \apjl, 615,
  L101, \dodoi{10.1086/426079}

\bibitem[{Barbary(2016)}]{Barbary2016}
Barbary, K. 2016, Journal of Open Source Software, 1, 58,
  \dodoi{10.21105/joss.00058}

\bibitem[{{Bell} {et~al.}(2004){Bell}, {Wolf}, {Meisenheimer}, {Rix}, {Borch},
  {Dye}, {Kleinheinrich}, {Wisotzki}, \& {McIntosh}}]{Bell2004}
{Bell}, E.~F., {Wolf}, C., {Meisenheimer}, K., {et~al.} 2004, \apj, 608, 752,
  \dodoi{10.1086/420778}

\bibitem[{Brammer {et~al.}(2008)Brammer, van Dokkum, \& Coppi}]{EAZY}
Brammer, G.~B., van Dokkum, P.~G., \& Coppi, P. 2008, The Astrophysical
  journal, 686, 1503

\bibitem[{{Brammer} {et~al.}(2011){Brammer}, {Whitaker}, {van Dokkum},
  {Marchesini}, {Franx}, {Kriek}, {Labb{\'e}}, {Lee}, {Muzzin}, {Quadri},
  {Rudnick}, \& {Williams}}]{Brammer2011}
{Brammer}, G.~B., {Whitaker}, K.~E., {van Dokkum}, P.~G., {et~al.} 2011, \apj,
  739, 24, \dodoi{10.1088/0004-637X/739/1/24}

\bibitem[{Brennan {et~al.}(2015)Brennan, Pandya, Somerville, Barro, Taylor,
  Wuyts, Bell, Dekel, Ferguson, McIntosh, Papovich, \& Primack}]{Brennan15}
Brennan, R., Pandya, V., Somerville, R.~S., {et~al.} 2015, Monthly notices of
  the Royal Astronomical Society, 451, 2933

\bibitem[{{Capak} {et~al.}(2015){Capak}, {Carilli}, {Jones}, {Casey},
  {Riechers}, {Sheth}, {Carollo}, {Ilbert}, {Karim}, {Lefevre}, {Lilly},
  {Scoville}, {Smolcic}, \& {Yan}}]{Capak2015}
{Capak}, P.~L., {Carilli}, C., {Jones}, G., {et~al.} 2015, \nat, 522, 455,
  \dodoi{10.1038/nature14500}

\bibitem[{Carnall {et~al.}(2018)Carnall, McLure, Dunlop, \& Dav{\'{e}
  }}]{Carnall18}
Carnall, A.~C., McLure, R.~J., Dunlop, J.~S., \& Dav{\'{e} }, R. 2018, Monthly
  Notices of the Royal Astronomical Society, 480, 4379,
  \dodoi{10.1093/mnras/sty2169}

\bibitem[{Carnall {et~al.}(2022)Carnall, McLeod, McLure, Dunlop, Begley,
  Cullen, Donnan, Hamadouche, Jewell, Jones, Pollock, \& Wild}]{Carnall2022}
Carnall, A.~C., McLeod, D.~J., McLure, R.~J., {et~al.} 2022, A first look at
  JWST CEERS: massive quiescent galaxies from 3 \&lt; z \&lt; 5,  arXiv,
  \dodoi{10.48550/ARXIV.2208.00986}

\bibitem[{Casey {et~al.}(2018)Casey, Hodge, Zavala, Spilker, da~Cunha, Staguhn,
  Finkelstein, \& Drew}]{Casey18}
Casey, C.~M., Hodge, J., Zavala, J.~A., {et~al.} 2018, The Astrophysical
  Journal, 862, 78, \dodoi{10.3847/1538-4357/aacd11}

\bibitem[{Cecchi {et~al.}(2019)Cecchi, Bolzonella, Cimatti, \&
  Girelli}]{Cecchi_2019}
Cecchi, R., Bolzonella, M., Cimatti, A., \& Girelli, G. 2019, The Astrophysical
  Journal, 880, L14, \dodoi{10.3847/2041-8213/ab2c80}

\bibitem[{{Choi} {et~al.}(2014){Choi}, {Rott}, \& {Itow}}]{Choi2014}
{Choi}, K., {Rott}, C., \& {Itow}, Y. 2014, \jcap, 2014, 049,
  \dodoi{10.1088/1475-7516/2014/05/049}

\bibitem[{{Conroy} \& {Gunn}(2010)}]{ConroyGunn10}
{Conroy}, C., \& {Gunn}, J.~E. 2010, \apj, 712, 833,
  \dodoi{10.1088/0004-637X/712/2/833}

\bibitem[{{Conroy} {et~al.}(2009){Conroy}, {Gunn}, \& {White}}]{Conroy09}
{Conroy}, C., {Gunn}, J.~E., \& {White}, M. 2009, \apj, 699, 486,
  \dodoi{10.1088/0004-637X/699/1/486}

\bibitem[{Davé {et~al.}(2019)Davé, Anglés-Alcázar, Narayanan, Li,
  Rafieferantsoa, \& Appleby}]{Dave2019}
Davé, R., Anglés-Alcázar, D., Narayanan, D., {et~al.} 2019, Monthly Notices
  of the Royal Astronomical Society, 486, 2827, \dodoi{10.1093/mnras/stz937}

\bibitem[{De~Lucia \& Blaizot(2007)}]{DeLucia07}
De~Lucia, G., \& Blaizot, J. 2007, Monthly notices of the Royal Astronomical
  Society, 375, 2

\bibitem[{{Dickey} {et~al.}(2021){Dickey}, {Starkenburg}, {Geha}, {Hahn},
  {Angl{\'e}s-Alc{\'a}zar}, {Choi}, {Dav{\'e}}, {Genel}, {Iyer}, {Maller},
  {Mandelker}, {Somerville}, \& {Yung}}]{Dickey2021}
{Dickey}, C.~M., {Starkenburg}, T.~K., {Geha}, M., {et~al.} 2021, \apj, 915,
  53, \dodoi{10.3847/1538-4357/abc014}

\bibitem[{Drew \& Casey(2022)}]{Drew_2022}
Drew, P.~M., \& Casey, C.~M. 2022, The Astrophysical Journal, 930, 142,
  \dodoi{10.3847/1538-4357/ac6270}

\bibitem[{{Faber} {et~al.}(2007){Faber}, {Willmer}, {Wolf}, {Koo}, {Weiner},
  {Newman}, {Im}, {Coil}, {Conroy}, {Cooper}, {Davis}, {Finkbeiner}, {Gerke},
  {Gebhardt}, {Groth}, {Guhathakurta}, {Harker}, {Kaiser}, {Kassin},
  {Kleinheinrich}, {Konidaris}, {Kron}, {Lin}, {Luppino}, {Madgwick},
  {Meisenheimer}, {Noeske}, {Phillips}, {Sarajedini}, {Schiavon}, {Simard},
  {Szalay}, {Vogt}, \& {Yan}}]{Faber2007}
{Faber}, S.~M., {Willmer}, C.~N.~A., {Wolf}, C., {et~al.} 2007, \apj, 665, 265,
  \dodoi{10.1086/519294}

\bibitem[{Fabian(2012)}]{Fabian_2012}
Fabian, A. 2012, Annual Review of Astronomy and Astrophysics, 50, 455,
  \dodoi{10.1146/annurev-astro-081811-125521}

\bibitem[{{Fang} {et~al.}(2012){Fang}, {Kong}, {Chen}, \& {Lin}}]{Fang2012}
{Fang}, G., {Kong}, X., {Chen}, Y., \& {Lin}, X. 2012, \apj, 751, 109,
  \dodoi{10.1088/0004-637X/751/2/109}

\bibitem[{Forrest {et~al.}(2020)Forrest, Annunziatella, Wilson, Marchesini,
  Muzzin, Cooper, Marsan, McConachie, Chan, Gomez, Kado-Fong, Barbera, Labbé,
  Lange-Vagle, Nantais, Nonino, Peña, Saracco, Stefanon, \& van~der
  Burg}]{Forrest20}
Forrest, B., Annunziatella, M., Wilson, G., {et~al.} 2020, Astrophysical
  journal. Letters, 890, L1

\bibitem[{{Franco} {et~al.}(2018){Franco}, {Elbaz}, {B{\'e}thermin},
  {Magnelli}, {Schreiber}, {Ciesla}, {Dickinson}, {Nagar}, {Silverman},
  {Daddi}, {Alexander}, {Wang}, {Pannella}, {Le Floc'h}, {Pope}, {Giavalisco},
  {Maury}, {Bournaud}, {Chary}, {Demarco}, {Ferguson}, {Finkelstein}, {Inami},
  {Iono}, {Juneau}, {Lagache}, {Leiton}, {Lin}, {Magdis}, {Messias},
  {Motohara}, {Mullaney}, {Okumura}, {Papovich}, {Pforr}, {Rujopakarn},
  {Sargent}, {Shu}, \& {Zhou}}]{Franco18}
{Franco}, M., {Elbaz}, D., {B{\'e}thermin}, M., {et~al.} 2018, \aap, 620, A152,
  \dodoi{10.1051/0004-6361/201832928}

\bibitem[{Fraser-McKelvie \& Cortese(2022)}]{Fraser_McKelvie_2022}
Fraser-McKelvie, A., \& Cortese, L. 2022, The Astrophysical Journal, 937, 117,
  \dodoi{10.3847/1538-4357/ac874d}

\bibitem[{Fujimoto {et~al.}(2017)Fujimoto, Ouchi, Shibuya, \&
  Nagai}]{Fujimoto_2017}
Fujimoto, S., Ouchi, M., Shibuya, T., \& Nagai, H. 2017, The Astrophysical
  Journal, 850, 83, \dodoi{10.3847/1538-4357/aa93e6}

\bibitem[{{Fujimoto} {et~al.}(2020){Fujimoto}, {Silverman}, {Bethermin},
  {Ginolfi}, {Jones}, {Le F{\`e}vre}, {Dessauges-Zavadsky}, {Rujopakarn},
  {Faisst}, {Fudamoto}, {Cassata}, {Morselli}, {Maiolino}, {Schaerer}, {Capak},
  {Yan}, {Vallini}, {Toft}, {Loiacono}, {Zamorani}, {Talia}, {Narayanan},
  {Hathi}, {Lemaux}, {Boquien}, {Amorin}, {Ibar}, {Koekemoer},
  {M{\'e}ndez-Hern{\'a}ndez}, {Bardelli}, {Vergani}, {Zucca}, {Romano}, \&
  {Cimatti}}]{Fujimoto20}
{Fujimoto}, S., {Silverman}, J.~D., {Bethermin}, M., {et~al.} 2020, \apj, 900,
  1, \dodoi{10.3847/1538-4357/ab94b3}

\bibitem[{Gebhardt {et~al.}(2021)Gebhardt, Cooper, Ciardullo, Acquaviva,
  Bender, Bowman, Castanheira, Dalton, Davis, de~Jong, DePoy, Devarakonda,
  Dongsheng, Drory, Fabricius, Farrow, Feldmeier, Finkelstein, Froning,
  Gawiser, Gronwall, Herold, Hill, Hopp, House, Janowiecki, Jarvis, Jeong,
  Jogee, Kakuma, Kelz, Kollatschny, Komatsu, Krumpe, Landriau, Liu, Niemeyer,
  MacQueen, Marshall, Mawatari, McLinden, Mukae, Nagaraj, Ono, Ouchi, Papovich,
  Sakai, Saito, Schneider, Schulze, Shanmugasundararaj, Shetrone, Sneden,
  Snigula, Steinmetz, Thomas, Thomas, Tuttle, Urrutia, Wisotzki, Wold, Zeimann,
  \& Zhang}]{Gebhardt_2021}
Gebhardt, K., Cooper, E.~M., Ciardullo, R., {et~al.} 2021, The Astrophysical
  Journal, 923, 217, \dodoi{10.3847/1538-4357/ac2e03}

\bibitem[{Glazebrook {et~al.}(2017)Glazebrook, Schreiber, Labbé, Nanayakkara,
  Kacprzak, Oesch, Papovich, Spitler, Straatman, Tran, \& Yuan}]{Glazebrook17}
Glazebrook, K., Schreiber, C., Labbé, I., {et~al.} 2017, Nature (London), 544,
  71

\bibitem[{{Hill} {et~al.}(2008){Hill}, {Gebhardt}, {Komatsu}, {Drory},
  {MacQueen}, {Adams}, {Blanc}, {Koehler}, {Rafal}, {Roth}, {Kelz}, {Gronwall},
  {Ciardullo}, \& {Schneider}}]{Hill08}
{Hill}, G.~J., {Gebhardt}, K., {Komatsu}, E., {et~al.} 2008, in Astronomical
  Society of the Pacific Conference Series, Vol. 399, Panoramic Views of Galaxy
  Formation and Evolution, ed. T.~{Kodama}, T.~{Yamada}, \& K.~{Aoki}, 115,
  \dodoi{10.48550/arXiv.0806.0183}

\bibitem[{{Ikarashi} {et~al.}(2017){Ikarashi}, {Caputi}, {Ohta}, {Ivison},
  {Lagos}, {Bisigello}, {Hatsukade}, {Aretxaga}, {Dunlop}, {Hughes}, {Iono},
  {Izumi}, {Kashikawa}, {Koyama}, {Kawabe}, {Kohno}, {Motohara}, {Nakanishi},
  {Tamura}, {Umehata}, {Wilson}, {Yabe}, \& {Yun}}]{Ikarashi2017}
{Ikarashi}, S., {Caputi}, K.~I., {Ohta}, K., {et~al.} 2017, \apjl, 849, L36,
  \dodoi{10.3847/2041-8213/aa9572}

\bibitem[{Kennicutt \& Evans(2012)}]{Kennicutt_2012}
Kennicutt, R.~C., \& Evans, N.~J. 2012, Annual Review of Astronomy and
  Astrophysics, 50, 531, \dodoi{10.1146/annurev-astro-081811-125610}

\bibitem[{King \& Pounds(2015)}]{King_2015}
King, A., \& Pounds, K. 2015, Annual Review of Astronomy and Astrophysics, 53,
  115, \dodoi{10.1146/annurev-astro-082214-122316}

\bibitem[{Kriek \& Conroy(2013)}]{Kriek_2013}
Kriek, M., \& Conroy, C. 2013, The Astrophysical Journal, 775, L16,
  \dodoi{10.1088/2041-8205/775/1/l16}

\bibitem[{{Kriek} {et~al.}(2006){Kriek}, {van Dokkum}, {Franx}, {Quadri},
  {Gawiser}, {Herrera}, {Illingworth}, {Labb{\'e}}, {Lira}, {Marchesini},
  {Rix}, {Rudnick}, {Taylor}, {Toft}, {Urry}, \& {Wuyts}}]{Kriek2006}
{Kriek}, M., {van Dokkum}, P.~G., {Franx}, M., {et~al.} 2006, \apjl, 649, L71,
  \dodoi{10.1086/508371}

\bibitem[{{Kroupa} \& {Weidner}(2003)}]{Kroupa2003}
{Kroupa}, P., \& {Weidner}, C. 2003, \apj, 598, 1076, \dodoi{10.1086/379105}

\bibitem[{{Lang} {et~al.}(2016){Lang}, {Hogg}, \& {Mykytyn}}]{Lang2016}
{Lang}, D., {Hogg}, D.~W., \& {Mykytyn}, D. 2016, {The Tractor: Probabilistic
  astronomical source detection and measurement}, Astrophysics Source Code
  Library, record ascl:1604.008.
\newblock \doeprint{1604.008}

\bibitem[{{Leung} {et~al.}(2023){Leung}, {Finkelstein}, {Weaver}, {Papovich},
  {Larson}, {Chworowsky}, {Ciardullo}, {Gawiser}, {Gronwall}, {Jogee},
  {Kawinwanichakij}, {Somerville}, {Wold}, \& {Yung}}]{Leung23}
{Leung}, G. C.~K., {Finkelstein}, S., {Weaver}, J., {et~al.} 2023, arXiv
  e-prints, arXiv:2301.00908, \dodoi{10.48550/arXiv.2301.00908}

\bibitem[{{Maiolino} {et~al.}(2015{\natexlab{a}}){Maiolino}, {Carniani},
  {Fontana}, {Vallini}, {Pentericci}, {Ferrara}, {Vanzella}, {Grazian},
  {Gallerani}, {Castellano}, {Cristiani}, {Brammer}, {Santini}, {Wagg}, \&
  {Williams}}]{Maiolino2015}
{Maiolino}, R., {Carniani}, S., {Fontana}, A., {et~al.} 2015{\natexlab{a}},
  \mnras, 452, 54, \dodoi{10.1093/mnras/stv1194}

\bibitem[{{Maiolino} {et~al.}(2015{\natexlab{b}}){Maiolino}, {Carniani},
  {Fontana}, {Vallini}, {Pentericci}, {Ferrara}, {Vanzella}, {Grazian},
  {Gallerani}, {Castellano}, {Cristiani}, {Brammer}, {Santini}, {Wagg}, \&
  {Williams}}]{Maiolino15}
---. 2015{\natexlab{b}}, \mnras, 452, 54, \dodoi{10.1093/mnras/stv1194}

\bibitem[{Marchesini \& van Dokkum(2007)}]{Marchesini07}
Marchesini, D., \& van Dokkum, P.~G. 2007, The Astrophysical journal, 663, L89

\bibitem[{Marrese {et~al.}(2018)Marrese, Marinoni, Fabrizio, \&
  Altavilla}]{gai21}
Marrese, P., Marinoni, S., Fabrizio, M., \& Altavilla, G. 2018, Astronomy \&
  Astrophysics, 621, \dodoi{10.1051/0004-6361/201834142}

\bibitem[{Merlin {et~al.}(2018)Merlin, Fontana, Castellano, Santini, Torelli,
  Boutsia, Wang, Grazian, Pentericci, Schreiber, Ciesla, McLure, Derriere,
  Dunlop, \& Elbaz}]{Merlin18}
Merlin, E., Fontana, A., Castellano, M., {et~al.} 2018, Monthly notices of the
  Royal Astronomical Society, 473, 2098

\bibitem[{Merlin {et~al.}(2019)Merlin, Fortuni, Torelli, Santini, Castellano,
  Fontana, Grazian, Pentericci, Pilo, \& Schmidt}]{Merlin2019}
Merlin, E., Fortuni, F., Torelli, M., {et~al.} 2019, Monthly Notices of the
  Royal Astronomical Society, 490, 3309, \dodoi{10.1093/mnras/stz2615}

\bibitem[{Muzzin {et~al.}(2013)Muzzin, Marchesini, Stefanon, Franx, McCracken,
  Milvang-Jensen, Dunlop, Fynbo, Brammer, Labbé, \& van Dokkum}]{Muzzin13}
Muzzin, A., Marchesini, D., Stefanon, M., {et~al.} 2013, The Astrophysical
  journal, 777, 18

\bibitem[{{Newman} {et~al.}(2018){Newman}, {Belli}, {Ellis}, \&
  {Patel}}]{Newman2018}
{Newman}, A.~B., {Belli}, S., {Ellis}, R.~S., \& {Patel}, S.~G. 2018, \apj,
  862, 125, \dodoi{10.3847/1538-4357/aacd4d}

\bibitem[{{Oke} \& {Gunn}(1983)}]{ABmag}
{Oke}, J.~B., \& {Gunn}, J.~E. 1983, \apj, 266, 713, \dodoi{10.1086/160817}

\bibitem[{Papovich {et~al.}(2016)Papovich, Shipley, Mehrtens, Lanham, Lacy,
  Ciardullo, Finkelstein, Bassett, Behroozi, Blanc, de~Jong, DePoy, Drory,
  Gawiser, Gebhardt, Gronwall, Hill, Hopp, Jogee, Kawinwanichakij, Marshall,
  McLinden, Cooper, Somerville, Steinmetz, Tran, Tuttle, Viero, Wechsler, \&
  Zeimann}]{Papovich_2016}
Papovich, C., Shipley, H.~V., Mehrtens, N., {et~al.} 2016, The Astrophysical
  Journal Supplement Series, 224, 28, \dodoi{10.3847/0067-0049/224/2/28}

\bibitem[{{Rinaldi} {et~al.}(2022){Rinaldi}, {Caputi}, {van Mierlo}, {Ashby},
  {Caminha}, \& {Iani}}]{Rinaldi2022}
{Rinaldi}, P., {Caputi}, K.~I., {van Mierlo}, S.~E., {et~al.} 2022, \apj, 930,
  128, \dodoi{10.3847/1538-4357/ac5d39}

\bibitem[{{Salim} {et~al.}(2018){Salim}, {Boquien}, \& {Lee}}]{Salim2018}
{Salim}, S., {Boquien}, M., \& {Lee}, J.~C. 2018, \apj, 859, 11,
  \dodoi{10.3847/1538-4357/aabf3c}

\bibitem[{Salmon {et~al.}(2015)Salmon, Papovich, Finkelstein, Tilvi, Finlator,
  Behroozi, Dahlen, Davé, Dekel, Dickinson, Ferguson, Giavalisco, Long, Lu,
  Mobasher, Reddy, Somerville, \& Wechsler}]{Salmon15}
Salmon, B., Papovich, C., Finkelstein, S.~L., {et~al.} 2015, The Astrophysical
  journal, 799, 183

\bibitem[{Sandles {et~al.}(2022)Sandles, Curtis-Lake, Charlot, Chevallard, \&
  Maiolino}]{Sandles_2022}
Sandles, L., Curtis-Lake, E., Charlot, S., Chevallard, J., \& Maiolino, R.
  2022, Monthly Notices of the Royal Astronomical Society, 515, 2951,
  \dodoi{10.1093/mnras/stac1999}

\bibitem[{{Santini} {et~al.}(2019){Santini}, {Merlin}, {Fontana}, {Magnelli},
  {Paris}, {Castellano}, {Grazian}, {Pentericci}, {Pilo}, \&
  {Torelli}}]{Santini19}
{Santini}, P., {Merlin}, E., {Fontana}, A., {et~al.} 2019, \mnras, 486, 560,
  \dodoi{10.1093/mnras/stz801}

\bibitem[{Schreiber {et~al.}(2017)Schreiber, Pannella, Leiton, Elbaz, Wang,
  Okumura, \& Labb{\'{e} }}]{Schreiber17}
Schreiber, C., Pannella, M., Leiton, R., {et~al.} 2017, Astronomy \&
  Astrophysics, 599, A134, \dodoi{10.1051/0004-6361/201629155}

\bibitem[{Schreiber {et~al.}(2018)Schreiber, Glazebrook, Nanayakkara, Kacprzak,
  Labbé, Oesch, Yuan, Tran, Papovich, Spitler, \& Straatman}]{Schreiber18}
Schreiber, C., Glazebrook, K., Nanayakkara, T., {et~al.} 2018, Astronomy and
  astrophysics (Berlin), 618, A85

\bibitem[{{Scoville} {et~al.}(2016){Scoville}, {Sheth}, {Aussel}, {Vanden
  Bout}, {Capak}, {Bongiorno}, {Casey}, {Murchikova}, {Koda},
  {{\'A}lvarez-M{\'a}rquez}, {Lee}, {Laigle}, {McCracken}, {Ilbert}, {Pope},
  {Sanders}, {Chu}, {Toft}, {Ivison}, \& {Manohar}}]{Scoville2016}
{Scoville}, N., {Sheth}, K., {Aussel}, H., {et~al.} 2016, \apj, 820, 83,
  \dodoi{10.3847/0004-637X/820/2/83}

\bibitem[{Somerville \& Davé(2015)}]{Somerville15_annualreview}
Somerville, R.~S., \& Davé, R. 2015, Annual review of astronomy and
  astrophysics, 53, 51

\bibitem[{Sommovigo {et~al.}(2022)Sommovigo, Ferrara, Carniani, Pallottini,
  Dayal, Pizzati, Ginolfi, Markov, \& Faisst}]{Sommovigo_2022}
Sommovigo, L., Ferrara, A., Carniani, S., {et~al.} 2022, Monthly Notices of the
  Royal Astronomical Society, 517, 5930, \dodoi{10.1093/mnras/stac2997}

\bibitem[{Speagle {et~al.}(2014)Speagle, Steinhardt, Capak, \&
  Silverman}]{Speagle2014}
Speagle, J.~S., Steinhardt, C.~L., Capak, P.~L., \& Silverman, J.~D. 2014, The
  Astrophysical Journal Supplement Series, 214, 15,
  \dodoi{10.1088/0067-0049/214/2/15}

\bibitem[{Spitler {et~al.}(2014)Spitler, Straatman, Labbé, Glazebrook, Tran,
  Kacprzak, Quadri, Papovich, Persson, van Dokkum, Allen, Kawinwanichakij,
  Kelson, McCarthy, Mehrtens, J.~Monson, Nanayakkara, Rees, Tilvi, \&
  Tomczak}]{Spitler14}
Spitler, L.~R., Straatman, C. M.~S., Labbé, I., {et~al.} 2014, Astrophysical
  journal. Letters, 787, L36

\bibitem[{{Springel} {et~al.}(2005){Springel}, {White}, {Jenkins}, {Frenk},
  {Yoshida}, {Gao}, {Navarro}, {Thacker}, {Croton}, {Helly}, {Peacock}, {Cole},
  {Thomas}, {Couchman}, {Evrard}, {Colberg}, \& {Pearce}}]{Springel2005a}
{Springel}, V., {White}, S. D.~M., {Jenkins}, A., {et~al.} 2005, \nat, 435,
  629, \dodoi{10.1038/nature03597}

\bibitem[{Stefanon {et~al.}(2013)Stefanon, Rudnick, Marchesini, Brammer, \&
  Whitaker}]{Stefanon13}
Stefanon, M., Rudnick, G.~H., Marchesini, D., Brammer, G.~B., \& Whitaker,
  K.~E. 2013, The Astrophysical journal, 768

\bibitem[{Stefanon {et~al.}(2015)Stefanon, Marchesini, Muzzin, Brammer, Dunlop,
  Franx, Fynbo, Labb{\'{e} }, Milvang-Jensen, \& van Dokkum}]{Stefanon_2015}
Stefanon, M., Marchesini, D., Muzzin, A., {et~al.} 2015, The Astrophysical
  Journal, 803, 11, \dodoi{10.1088/0004-637x/803/1/11}

\bibitem[{{Stevans} {et~al.}(2021){Stevans}, {Finkelstein}, {Kawinwanichakij},
  {Wold}, {Papovich}, {Somerville}, {Yung}, {Sherman}, {Ciardullo}, {Dav{\'e}},
  {Florez}, {Gronwall}, \& {Jogee}}]{Stevans21}
{Stevans}, M.~L., {Finkelstein}, S.~L., {Kawinwanichakij}, L., {et~al.} 2021,
  \apj, 921, 58, \dodoi{10.3847/1538-4357/ac0cf6}

\bibitem[{Straatman {et~al.}(2015)Straatman, Labb{\'{e} }, Spitler, Glazebrook,
  Tomczak, Allen, Brammer, Cowley, van Dokkum, Kacprzak, Kawinwanichakij,
  Mehrtens, Nanayakkara, Papovich, Persson, Quadri, Rees, Tilvi, Tran, \&
  Whitaker}]{Straatman_2015}
Straatman, C. M.~S., Labb{\'{e} }, I., Spitler, L.~R., {et~al.} 2015, The
  Astrophysical Journal, 808, L29, \dodoi{10.1088/2041-8205/808/1/l29}

\bibitem[{Valentino {et~al.}(2020)Valentino, Tanaka, Davidzon, Toft,
  Gómez-Guijarro, Stockmann, Onodera, Brammer, Ceverino, Faisst, Gallazzi,
  Hayward, Ilbert, Kubo, Magdis, Selsing, Shimakawa, Sparre, Steinhardt, Yabe,
  \& Zabl}]{ValentinoFrancesco2020QG1B}
Valentino, F., Tanaka, M., Davidzon, I., {et~al.} 2020, The Astrophysical
  journal, 889, 93

\bibitem[{Viero {et~al.}(2014)Viero, Asboth, Roseboom, Moncelsi, Marsden,
  Cooper, Zemcov, Addison, Baker, Beelen, Bock, Bridge, Conley, Devlin,
  Dor{\'{e} }, Farrah, Finkelstein, Font-Ribera, Geach, Gebhardt, Gill, Glenn,
  Hajian, Halpern, Jogee, Kurczynski, Lapi, Negrello, Oliver, Papovich, Quadri,
  Ross, Scott, Schulz, Somerville, Spergel, Vieira, Wang, \&
  Wechsler}]{Viero_2014}
Viero, M.~P., Asboth, V., Roseboom, I.~G., {et~al.} 2014, The Astrophysical
  Journal Supplement Series, 210, 22, \dodoi{10.1088/0067-0049/210/2/22}

\bibitem[{{Wold} {et~al.}(2019){Wold}, {Kawinwanichakij}, {Stevans},
  {Finkelstein}, {Papovich}, {Devarakonda}, {Ciardullo}, {Feldmeier}, {Florez},
  {Gawiser}, {Gronwall}, {Jogee}, {Marshall}, {Sherman}, {Shipley},
  {Somerville}, {Valdes}, \& {Zeimann}}]{Wold19}
{Wold}, I. G.~B., {Kawinwanichakij}, L., {Stevans}, M.~L., {et~al.} 2019,
  \apjs, 240, 5, \dodoi{10.3847/1538-4365/aaee85}

\end{thebibliography}
\bibliographystyle{aasjournal}


\appendix \label{sec:appendix}
\section{\textsc{Bagpipes} Fits}

\begin{figure*}[h]
\centering
\includegraphics[width=0.95\linewidth]{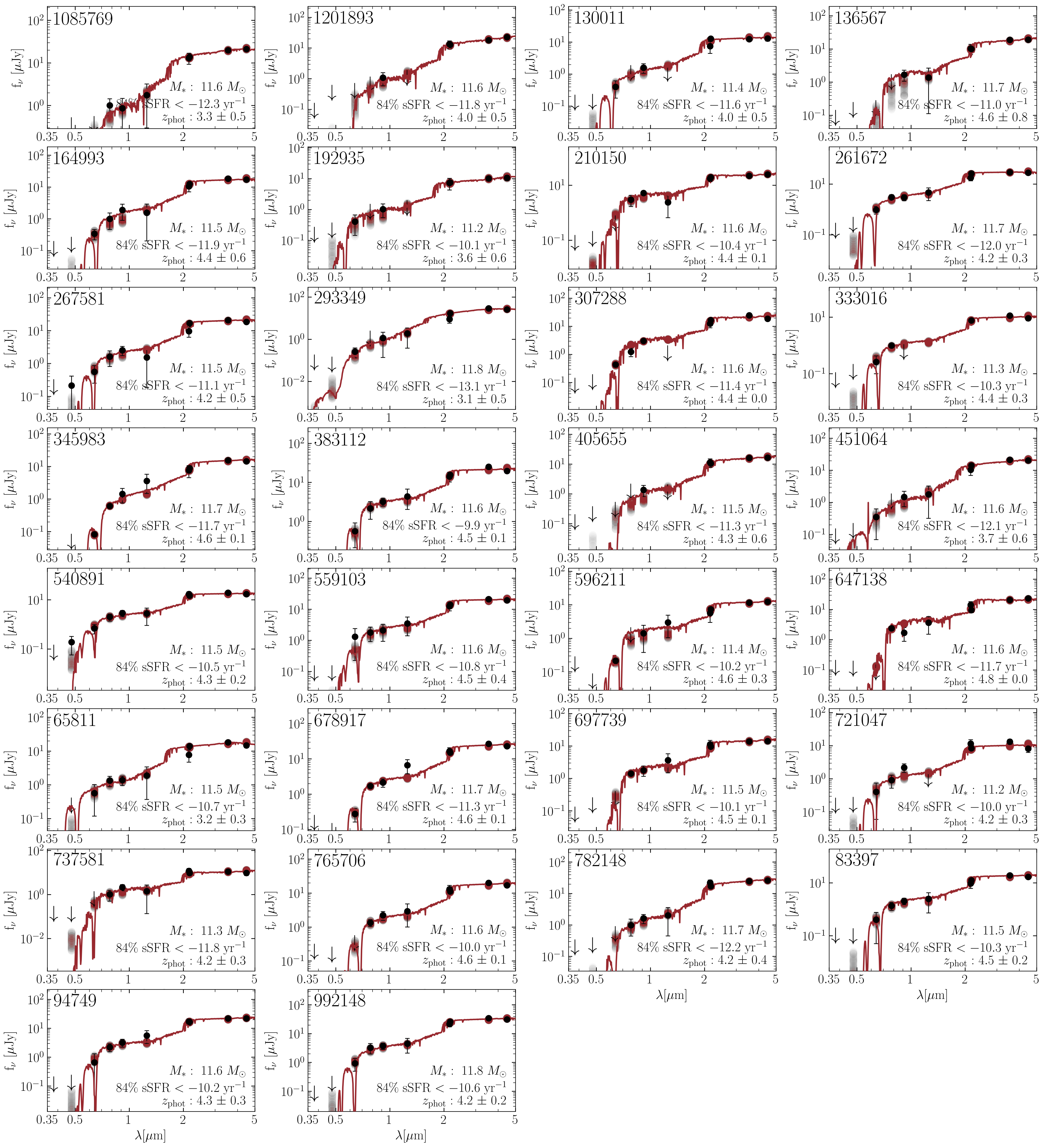}
\caption{SEDs for selected massive high redshift quiescent galaxies, which were not detected by ALMA. The best-fit SED is in red, photometry are denoted by the black points and the red shaded points are 1$\sigma$ posteriors on the photometry.\label{fig:BP_SEDs_noALMA}}
\end{figure*}

\begin{figure*}[h]
\centering
\includegraphics[width=0.9\linewidth]{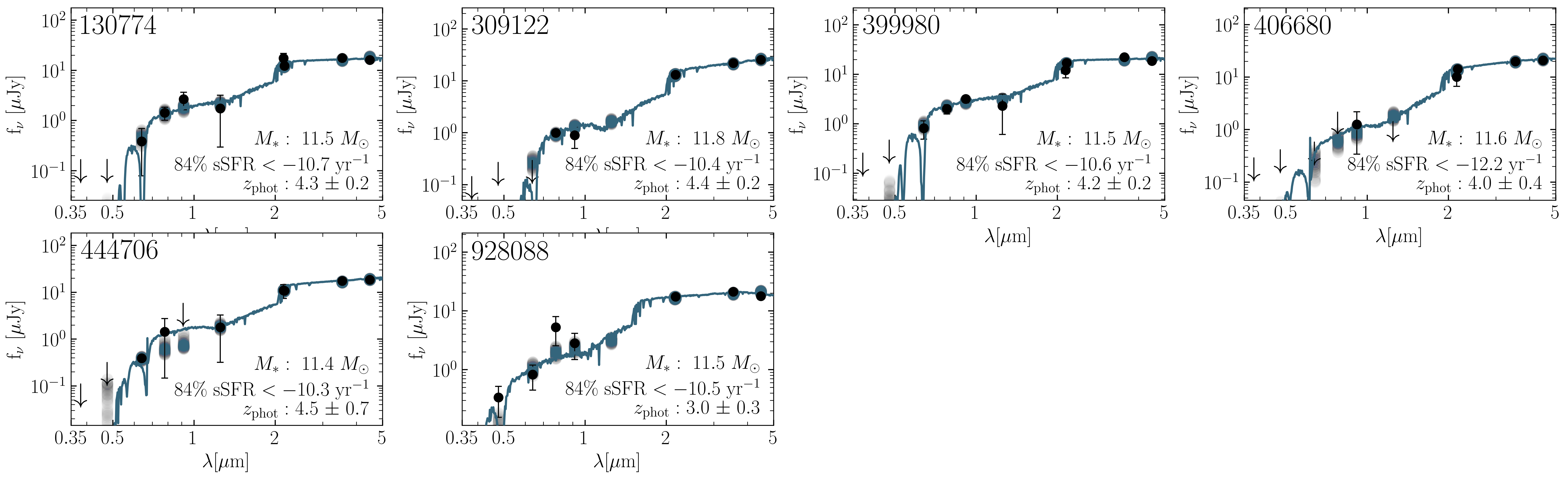}
\caption{SEDs for candidate massive high redshift quiescent galaxies which were removed based flux measurements in ALMA 1.1mm. While in the optical-IR, these sources seem quiescent based on the best-fitting SED, the measured dust-obscured star-formation rate from 1.1mm flux for these sources place them at $<2 \sigma$ below the SFMS at $z\sim 4$. The best-fit SED is in blue, photometry are denoted by the black points and the blue shaded points are 1$\sigma$ posteriors on the photometry.\label{fig:BP_SEDs_ALMA}}
\end{figure*}

\section{ALMA observations}\label{sec:not_selected}
\begin{figure*}[h]
\centering
\includegraphics[width=0.9\linewidth]{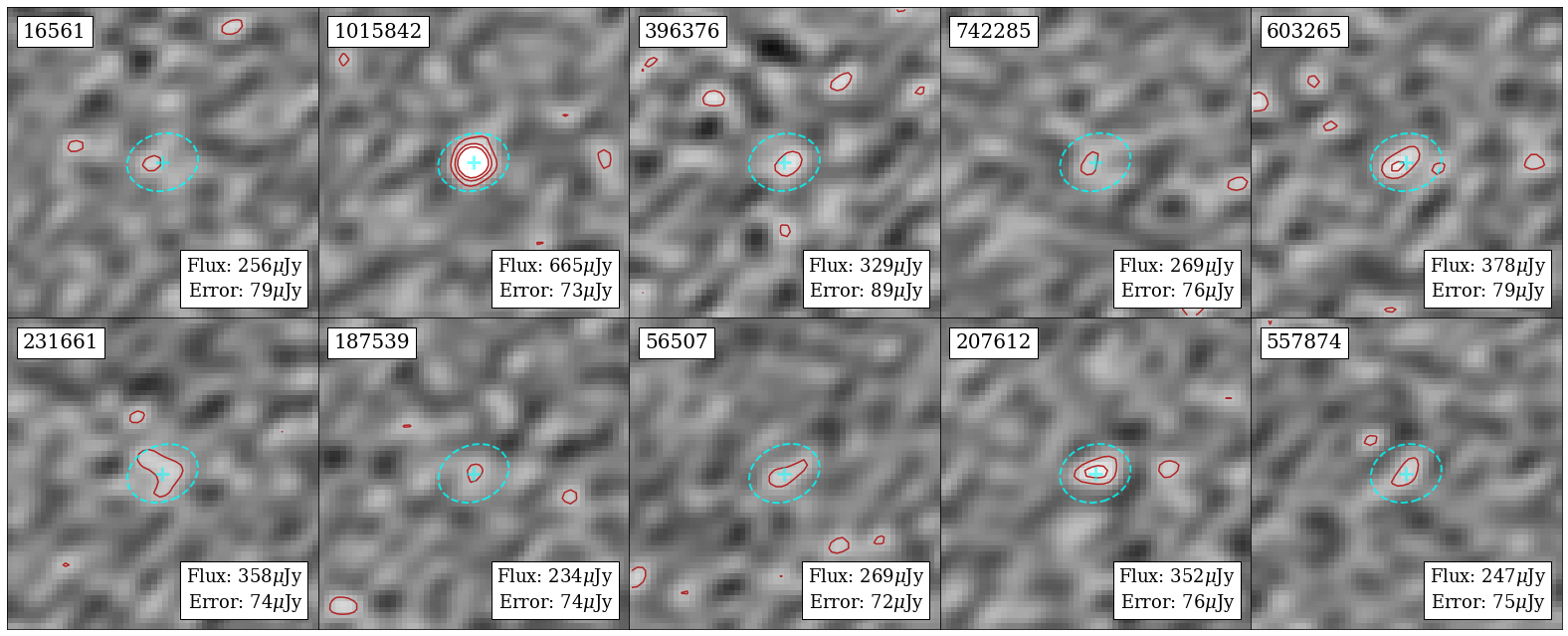}
\caption{ALMA observations for 10 sources which did not satisfy the selection criteria presented in this paper and have significant flux detected in 1.1mm. Each stamp is 10"$\times$10". These observations indicate the presence of dust-obscured star-formation. Some sources have an apparent offset between the source position, indicated by the cyan cross in the center of each stamp, and the ALMA flux. This offset between stellar emission and dust radiation has been observed in several other similar studies \citep[][e.g.]{Maiolino15, Franco18, Fujimoto20}, the ALMA flux detection here has been determined to be from the targeted source and within the $K$-band PSF.\label{fig:notselected_d}}
\end{figure*}

\begin{figure*}[h]
\centering
\includegraphics[width=0.9\linewidth]{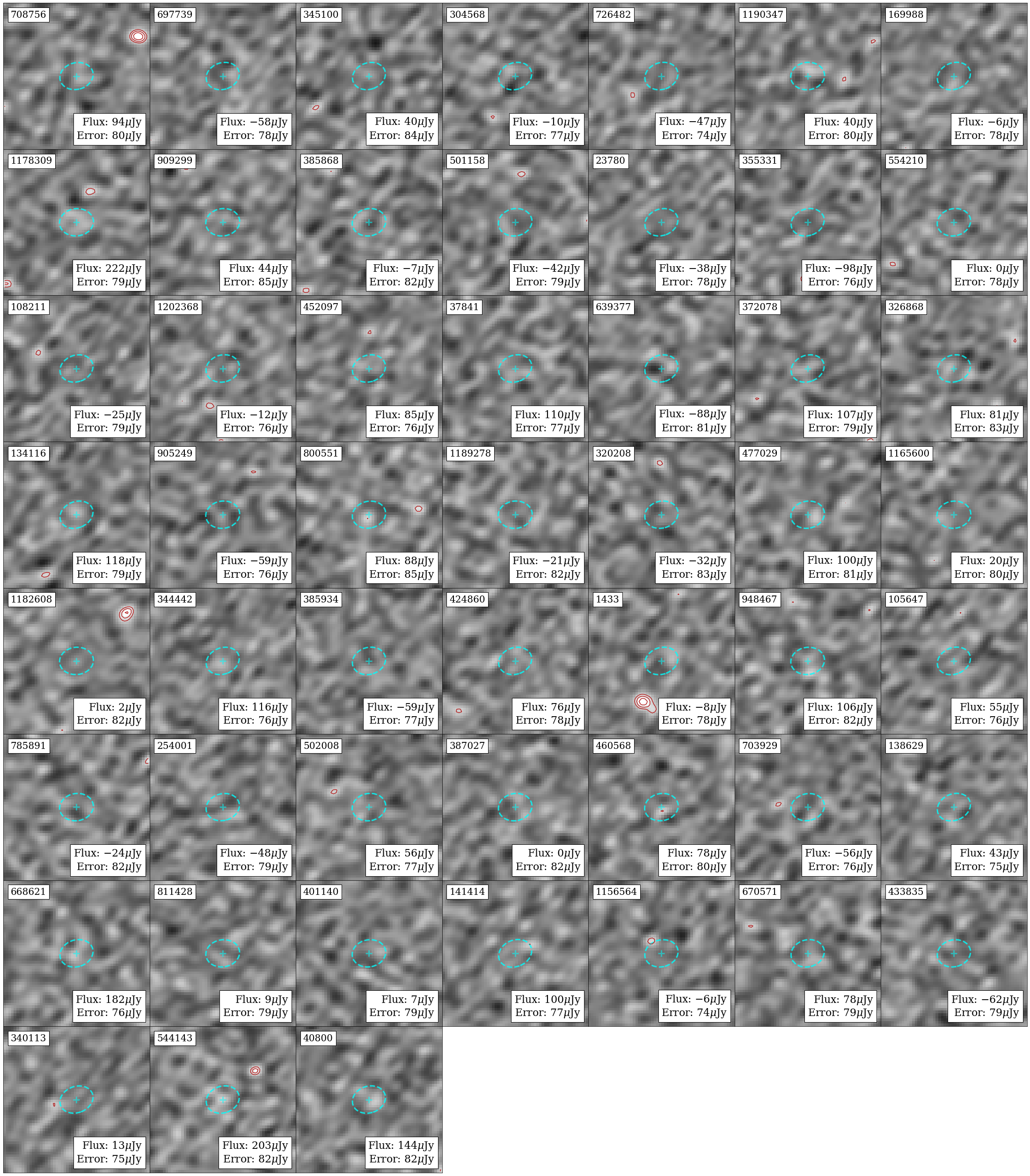}
\caption{ALMA observations for 52 sources which were not considered high redshift and quiescent. Each stamp is 10"$\times$10". These sources do not have significant flux in 1.1 mm. We stress that while we do not include these sources in our final sample, due to the conservative nature of our selection, discussed in Section \ref{sec:selecting}, many of these sources do still have sSFRs well below the SFMS.  \label{fig:notselected_nd}}
\end{figure*}

Here we present ALMA observations for the 62 sources which did not satisfy our selection criteria to be considered high redshift and quiescent. 52 of these sources had no significant detection in ALMA 1.1mm imaging, shown in Figure \ref{fig:notselected_nd}. While the other 10 sources were determined to have a $>3 \sigma$ detection in 1.1$\mu$m, these are shown in Figure \ref{fig:notselected_d}. 

\clearpage

\end{document}